\documentclass[notitlepage,nofootinbib,
twocolumn,
longbibliography,superscriptaddress,amsmath,amssymb]{revtex4-1}

\usepackage{graphicx}  
\usepackage{dcolumn}   
\usepackage{bm}        
\usepackage{verbatim}   
\usepackage{bbold}
\usepackage{color}
\usepackage{xcolor}
\usepackage{amsthm}
\usepackage{ulem}

\def\D{\Delta}

\def\s{\sigma}

\def\e{\epsilon}
\def\be{\begin{equation}}
\def\ee{\end{equation}}
\def\bea{\begin{eqnarray}}
\def\eea{\end{eqnarray}}

\def\>{\rangle}
\def\<{\langle}

\newcommand{\dbar}{\mathchar '26 \mkern-11mu d}

\theoremstyle{definition}

\theoremstyle{remark}

\begin{document}

\title{Modeling the pseudogap metallic state in cuprates: \\quantum disordered  pair density wave}

\author{Zhehao Dai}
\author{T. Senthil}
\author{Patrick A. Lee}
\affiliation{
Department of Physics, Massachusetts Institute of Technology, Cambridge, MA, USA
}

\date{June 4, 2019}

\begin{abstract}
We present a way to quantum-disorder a pair density wave, and propose it to be a candidate of the effective low-energy description of the pseudogap metal which may reveal itself in a sufficiently high magnetic field that suppresses the d-wave pairing. The ground state we construct is a small-pocket Fermi liquid with a bosonic Mott insulator in the density-wave-enlarged unit cell. At low energy, the charge density is mainly carried by charge $2e$ bosons, which develop a small insulating gap. As an intermediate step, we discuss the quantum disordering of a fully gapped superconductor and its excitation spectrum. In order to illustrate the concepts we introduce, we introduce a simplified 1D model which we solve numerically. We discuss a number of experimental consequences.  The interplay between the electron and the small-gap boson results in a step function background in the  electron spectral function which may be consistent with existing ARPES data. Optical excitation across the boson gap can explain the onset and the magnitude of the mid infra-red absorption reported long ago.
\end{abstract}

\maketitle

\section{Introduction}

The pseudogap phase occupies a large region in the phase diagram of underdoped cuprates, up to a  temperature much higher than the superconducting $T_c$. It marks the crossover from the lightly doped Mott insulator regime, where the electronic state is well described by hole pockets holding $p$ carriers, where $p$ is the density of doped holes, to the Fermi liquid state where a large Fermi surface holding $1+p$ holes is clearly observed. The energy gap itself starts out being very large (several hundreds meV) near the Mott insulator, but it persists in the anti-nodal region (the vicinity of $(0,\pi)$ in the Brillouin zone) for intermediate doping, where it co-exists with superconductivity. In this paper the term pseudogap phase refers to this intermediate doping regime, roughly in the range between $p$ = 0.08 to 0.19 in YBCO, where the pseudogap itself is about 80 meV or less. This regime has been under intense study and is commonly considered to be a central puzzle in the cuprate high-Tc problem~\cite{keimer2015quantum}.

Recent data show that this phase is marked by a thermodynamical phase transition~\cite{sato2017thermodynamic,zhao2017global,shekhter2013bounding}, but the nature of the order is controversial. Proposals range from a primary but hard to detect order such as intra-cell orbital current~\cite{PhysRevB.55.14554}, to composite order such as nematicity with the primary order parameter being disordered~\cite{fernandes2019intertwined}. At lower temperature, static charge order appears,~\cite{JulienNature477191wu2011magnetic,wu2013emergence,ZX1science350949gerber2015three,changNatureComm72016magnetic,ZX2PNAS11314647jang2016ideal} but there is now general agreement that the anti-nodal gap is not caused by the charge order.
A key part of the phenomenology is the discovery of the "Fermi arc" near the nodal regime (the vicinity of $(\pi/2,\pi/2)$ in the Brillouin zone), out of which the d wave pairing gap develops. There is a a large literature on the origin of the anti-nodal gap and the Fermi arc, ranging from fluctuating anti-ferromagnet~\cite{PhysRevB.94.205117}, spiral  spin density wave~\cite{PhysRevLett.117.187001},  Umklapp scattering of pairs of electrons~\cite{robinson2019anomalies,PhysRevB.73.174501,PhysRevB.95.201112} to fluctuating superconductivity of some kind~\cite{emery1995importance,PhysRevX.4.031017}. Yet another important phenomenology is that for doping near 1/8, the superconductivity is suppressed by an unexpectedly small magnetic field of about 20T ~\cite{chang2012decrease,grissonnanche2014direct,Julien2arXivzhou2017spin}. At higher field quantum oscillations identified with very small  electron like pockets have been observed~\cite{doiron2007quantum,sebastian2011quantum,ramshaw2011angle}.

It has proven to be extremely challenging to develop a theoretical picture to describe this rich and unexpected set of phenomena. Theoretical efforts can be roughly divided into two classes. The first involves microscopic theories that start with a model Hamiltonian such as the Hubbard model and attempt to solve for the low energy properties. Due to the complexity of the strong correlation problem, progress along this line has been made mainly with numerical methods. Approximate methods such as cluster DMFT (dynamical mean field theory) have shown that the Hubbard model indeed exhibit a  phase  where anti-nodal gap and near nodal gapless  carries co-exists and that this state undergoes d wave paring at low temperature~\cite{PhysRevLett.110.216405}. Other methods such as DMRG (density matrix renormalization group)~\cite{PhysRevB.98.140505}, Monte Carlo studies of projected wavefunctions~\cite{himeda2002},  exact diagonalization~\cite{PhysRevLett.113.046402} and other cluster embedding methods~\cite{zheng2017stripe}, provide information mainly on the ground state and its competitors. There appear to be a general consensus that while the d wave superconductor is a favored ground state, there exists a large variety of states that are very close in energy~\cite{zheng2017stripe}. These include various density waves with energy that is surprisingly insensitive to the period. 

A second line of attack is to do phenomenological theory. Here one postulate the existence of certain state or certain dominant order, and attempt to explain as much of the pseudogap phenomenology as possible based on the postulate. In view of the large variety of observations, even this is a highly nontrivial task. In this paper we follow this second line of attack. For reasons that will be explained below, our postulate is that the origin of the anti-node gap is from a quantum disordered (fluctuating in space and time) pair density wave (PDW). Furthermore in this paper we will mostly focus on a description of the zero temperature quantum state that emerges once the d wave pairing state is destroyed by a magnetic field. The focus on a quantum state and its low lying excitations allow us to make sharp statements. On the other hand, we will also make some qualitative predictions at zero field and finite temperature, taking advantage of the fact that the pseudogap scale $T_{PG}\gg T_{c}$.
 
Historically a popular starting point is to assume that in the underdoped region, due to the small superfluid stiffness, phase fluctuations greatly suppress the superconducting $T_c$ and the pseudogap is due to a large pairing amplitude that survives up to high temperature~\cite{emery1995importance}. However a d wave pairing gives rise to nodal points and it is not easy to obtain Fermi arcs in this scenario. Recently one of us ~\cite{PhysRevX.4.031017} has proposed that a different kind of pairing called pair density wave (PDW) where the Cooper pair carries a finite momentum is the ``mother state" that lies behind the pseudogap phenomenology. The PDW in cuprate has a rather long history, starting with the work of Himeda et al.~\cite{himeda2002} who proposed that associated with the stripes observed in the LBCO family, the pair order parameter has a sign change across  the nodes of the period 4 charge order, resulting in a period 8 PDW.  A great deal of experimental and theoretical progress have been made since that time, lending strong  support for this picture in the LSBO/LSCO  family, as recently reviewed by Agterberg et al.~\cite{agterberg2019review}.  However, an important distinction made in Ref.~\cite{PhysRevX.4.031017} and one which will play a key role in the present paper, is that the PDW is assumed to be bi-directional, in contrast to the uni-directional that has been discussed in connection to stripe physics in LBCO.
 
Compared with a fluctuating d wave superconductor, the proposal of a fluctuating PDW has a number of advantages. The PDW gaps out only the anti-nodal region of the Fermi surface and naturally leads to gapless excitations that resemble the Fermi arc. (Strictly speaking the arc is part of a closed loop made up of quasi-particles, but the backside of the loop is mainly hole-like and not visible in ARPES.) As explained in detail in Ref.~\cite{PhysRevX.4.031017} the single particle spectrum also shows many of the unusual features observed in a detailed study of single layer Bi-2201, where the superconducting $T_c$ is low and the pseudogap spectrum is accessible~\cite{he2011single,hashimoto2014energy}. An additional advantage is that the CDW appears naturally as a composite order 
 with the PDW as primary, thus making it unnecessary to postulate the CDW order as a separate instability. Since this point will play an essential role in the current paper, we give a more detailed explanation here.
 
Our construction assumes bi-directional PDWs with wave-vectors $P_x$ and $P_y$ which are characterized by four PDW order parameters, $\D_{P\hat{x}}$, $\D_{-P\hat{x}}$, $\D_{P\hat{y}}$, and $\D_{-P\hat{y}}$, with  equal amplitudes. One notices immediately that a term in the Landau free energy that couples linearly to density wave order is allowed by symmetry: $\rho_{2P\hat{x}}\D_{P\hat{x}}\D^*_{-P\hat{x}}$ This means that an ordered PDW with wave-vector $P\hat{x}$ necessarily induces a secondary order of CDW at wave vector $2P\hat{x}$. Perhaps less obvious is the notion that even if the primary order is fluctuating in space and time, a static and long range CDW order can also be induced, under the right circumstances. Consider the case when the phases of $\D_{P\hat{x}}$ and $\D_{-P\hat{x}}$ are wildly fluctuating but the relative phase between them is not. The linear coupling term will induce long range CDW order. Whether this happens or not depends on detailed choices of model parameters and this kind of phase diagram has been explicitly demonstrated in special cases~\cite{agterberg2019review,agterberg2008dislocations,berg2009charge}. This kind of possibility has been given the name vestigial order in a related disorder-driven case~\cite{nie2014quenched}, but we will continue to use the term composite order in this paper.

For the bi-directional PDW, a second possibility exists, ie CDW at wave-vector $P\hat{x}+P\hat{y}$ may be induced by the term:
$\rho_{P\hat{x}+P\hat{y}}(\D_{P\hat{x}}\D^*_{-P\hat{y}}+ \D^*_{-P\hat{x}}\D_{P\hat{y}})$. However, such a CDW has not been seen experimentally. Fortunately, Ref.~\cite{agterberg2008dislocations} has provided an explanation. They pointed out that there is another term that couples to an orbital magnetization density wave (MDW) which takes the form: $M_{P\hat{x}+P\hat{y}}\cdot i(\D_{P\hat{x}}\D^*_{-P\hat{y}} -  \D^*_{-P\hat{x}}\D_{P\hat{y}})$. The MDW involves orbital current at a finite wave-vector which produces an orbital magnetization. Note that the magnetization comes from orbital current and not spin, because the PDW order is a total spin singlet and will not couple to the spin degree of freedom in the absence of spin-orbit coupling. It turns out the  two terms inside the parenthesis in these Landau free energy terms either add or cancel each other, depending on their relative phase. Since CDW at $P\hat{x}+P\hat{y}$ is not observed, we assume the PDW order parameters have the phases that $\D_{P\hat{x}}\D^*_{-P\hat{y}}= - \D^*_{-P\hat{x}}\D_{P\hat{y}}$, such that the contribution to CDW cancels out, but MDW is stabilized at this wave-vector. The MDW may be detectable by neutron scattering, as will be discussed in Sec.~\ref{subsection:Symmetry breaking in the pseudopgap phase.}.

After this detour, we return to describe our basic postulate for the pseudogap phase. We assume the existence of a robust PDW amplitude over a large part of the doping and temperature range that is associate with the pseudogap. We assume that order is suppressed by quantum phase fluctuations. We further assume that CDW at twice the PDW wave-vector and MDW at 
$P\hat{x}\pm P\hat{y}$ are generated as long range ordered composite orders. In anticipation of what follows, we emphasize that the MDW (as a composite order) plays no role in producing the anti-nodal gap, but it will play an important role in determining the size of the reduced Brillouin zone (BZ) due to the increase periodicity. 

In support of this postulate, we have been greatly encouraged by a recent discovery of CDW order at half the conventional wave-vector in the vicinity of the vortex core in Bi-2212~\cite{edkins2019magnetic}.
In the presence of uniform d-wave superconducting order, a similar Landau argument predicted that a composite CDW may exist at wave-vector $P_x$ and $P_y$, exactly half that of the previously observed CDW wave-vector.

There have been two papers that provide theoretical interpretation of the observed data. The paper by Wang et al. (Ref.~\cite{PhysRevB.97.174510}) assumes that the PDW is a competing order and they constructed a sigma model with the uniform d wave and the PDW as  components. In this picture, the d wave order is suppressed near the vortex core in a region called the vortex halo and the PDW becomes stable. This picture produces a short range but static PDW in the vortex halo, and hence explain the short range and static CDW at $P_x$ and $P_y$.
The paper by Dai et al. (Ref.~\cite{PhysRevB.97.174511}) takes a somewhat different point of view. They assume, as we do here, that the PDW is always fluctuating, ie. it is not the ordered ground state even if the d wave superconductor is somehow suppressed. In order to create a static PDW near the vortex core, they assumed that the phase of the PDW is pinned by the rapid winding of the d wave pairing phase near the vortex core. This is possible if the wave-vector of the PDW matches the local phase gradient of the d wave order near the vortex core. The size of the vortex halo is then determined by the correlation length of the fluctuating PDW. As far as the STM data are concerned, both points of views seem to work, but we believe the sigma model picture will run into some difficulty if we ask the question of what happens when the vortex  halos overlap. Clearly d wave superconductivity will be destroyed. The relatively large size of the vortex halo will explain why the destruction of d wave superconductivity occurs at an unexpectedly low magnetic field. However, in the sigma model picture, it is difficult to see how one can avoid the conclusion that the resulting state is a long range ordered PDW and therefore a genuine superconductor. In contrast, in the fluctuating PDW scenario, the static PDW will be liberated and becomes freely fluctuating again once the pinning due to the d wave phase winding disappears for H greater than $\text{H}_{c2}$ and d wave pairing is killed.  By following this logic, the fluctuating PDW point of view advocated in Ref.~\cite{PhysRevX.4.031017,PhysRevB.97.174511} leads us naturally to the following question. What is the nature of the state that appears when H exceeds $\text{H}_{c2}$ when the vortex halos overlap? This will be a state with strong PDW amplitude whose long range order is destroyed by strong quantum phase fluctuations. We note that at least one experimental group~\cite{yu2016magnetic} has postulated that a fluctuating pairing state of some kind (which they call a quantum vortex liquid) permeates over a large region of the H-T plane. The goal of this paper is to clarify the nature of this state and expose as much of its physical properties as possible. A number of questions immediately come to mind:

1. Is this a metallic state? If so, is it smoothly connected to a conventional metal? Do we need to appeal to exotic concepts such as electron fractionalization and topological order to describe this state? If there is a Fermi surface, does it obey Luttinger's theorem?

2. Does this state have small electron pockets that is consistent with those observed in quantum oscillation experiments?

3. What is the nature of the excitation spectrum? If it is born of a PDW superconductor, the excitations started out as Bogoliubov quasi-particles, ie, superposition of particle and hole. Do they evolve to the excitations with fixed integer charges and if so, how does this evolution take place?

4. Why is there no sign of superconducting fluctuations in transport data in this high field low temperature regime. Are there other signs of the superconducting fluctuations? Are there other data that can be explained by this point of view that are difficult to explain otherwise? 

Producing answers to these questions seem at first glance to be a daunting task. 
The interplay between the phase fluctuation of superconductivity and gapless electron modes is usually difficult to deal with. Previous theoretical discussions on quantum disordering a zero-momentum d wave superconductor, which has gapless electron nodes, lead to models with fractional degrees of freedoms, and the discussions are yet to be settled (Ref.~\cite{PhysRevB.60.1654,PhysRevB.66.054535}). In our case the PDW has gaps only near the anti-nodes and a gapless region exists in the form of Fermi arcs. The gapless excitations seem to make the task even harder. However, it turns out that the composite order in the form of CDW comes to our rescue. The CDW connects the Fermi arcs and produces electron pockets, in the way proposed by Harrison and Sebastian~\cite{PhysRevLett.106.226402}.
The advantage here is that while Harrison and Sebastian had to postulate an anti-nodal gap of unknown origin, our state has the gap already in place. 
An important bonus of this picture is that in the new reduced BZ, the only gapless exciations are those of the electron pockets and these are naturally decoupled from the Bogoliubov-like quasi-particles that are associated with the PDW pairing. This is the key insight that allows us to make progress on this long standing problem.

In the discussion above, we have assumed the anti-nodal PDW gap of Bogoliubov excitations persists as an electron gap when PDW is disordered. It is known for a long time that a pairing gap of electron can survive even if the phase coherence of superconductivity is destroyed~\cite{PhysRevB.39.2756,PhysRevLett.47.1556,bouadim2011single}. The problem is usually mapped to Bose fluid with charge 2e electron pairs. After disordering the phase coherence, we get an insulator of pairs. We use this picture of bosonic superfluid-insulator transition to explain the gapped anti-nodal electron spectrum in the PDW-disordered ground state. Note that this bosonic description is usually 
not useful for  weakly interacting clean superconductors, because the size of the pair is much larger than the distance between pairs. However,  in our case, as discussed in later sections, the size of the pair is comparable to both the distance between pairs and the size of the enlarged unit cell due to composite orders; therefore we are in an intermediate regime and the insights from the BEC limit may serve as a guide. In particular, Coulomb repulsion between neighboring pairs may drive us to the Mott insulator state.  In this paper we provide a theoretical analysis of the evolution of the electron spectrum across the PDW disordering transition. The proposed spectral function on the PDW-disordered side can be compared directly with ARPES data on cuprates. 

The spectral similarity between the ordered PDW state and the pseudogap phase has been pointed out before, but the conceptual question about how the spectral feature of PDW, which is a superconducting order, can be used to explain the spectrum of the metallic pseudogap state is never discussed carefully. We answer this question in the current paper, and point out new spectral features due to the fluctuation of PDW. To the best of our knowledge, the momentum-dependent electron spectrum has not been discussed even in the simpler case of  s wave superconductor to insulator transition; therefore our analysis should be of broader  interests.

We proceed as follows. In Sec.~\ref{Sec: big picture} we provide the physical picture of the fluctuating PDW ground state. In Sec.~\ref{Sec: Constructing the fluctuating PDW ground state}, we present detailed analysis of the ground state, focusing on the evolution of electron spectral function. In Sec.~\ref{Sec: broader aspects experiments} we discuss the physical implications of the constructed ground state, compare it with existing experiments, and discuss new theoretical predictions.

\section{Fluctuating PDW and bosonic Mott insulator}\label{Sec: big picture}

In this section, we describe a way to quantum disorder the PDW to arrive at the desired pseudogap ground state. As discussed in the introduction, the PDW ansatz we consider has the nice property that the gapless nodal electrons and the gapped anti-nodal electrons form separate bands in the folded B.Z. Thus, we can treat the effect of fluctuating PDW separately for nodal electrons and anti-nodal electrons. 

For anti-nodal electrons, the central puzzle is whether the anti-nodal gap persists as an electron gap when PDW is disordered, and if so, how to understand the gapped electron spectrum when PDW is disordered. Since the same problem already exists in the simpler case of zero-momentum s-wave pairing, we first discuss the physics in this simplified situation.

In general, there are two possibilities when superconductivity is quantum disordered at zero temperature. If the pairing is weak, the electron gap may vanish immediately when superconductivity is disordered. Alternatively, if the pairing amplitude is large, and the superconductivity is destroyed by the phase fluctuation of its order parameter, we can destroy the superconducting long range order without closing the single-electron gap. The latter case is often treated by a bosonic theory~\cite{PhysRevB.39.2756,PhysRevLett.47.1556}.

We can understand the persistence of the electron gap in the BEC limit first. In this limit, electrons form tightly bound pairs, and the single-electron gap is just the binding energy of the pair, which is well-defined even for a single pair (like a molecule), therefore independent of whether the pairs condense or not. When the condensate is destroyed by phase fluctuations, the bosonic pairs may form a Mott insulator (if its density happens to be commensurate with the underlying lattice), a Wigner crystal that further breaks translation symmetries, or simply pinned by disorders. No matter which quantum state the bosonic pairs go to, the single-electron gap always persists when the superconductivity is disordered. 

In this work, we extrapolate from the BEC limit to the intermediate pairing regime, where the pairing amplitude is comparable or smaller than the Fermi energy but not too small. When we gradually reduce the pairing amplitude from the BEC limit to the intermediate pairing regime, by continuity, we expect the same transition from the superconductor to the paired insulator still exists, and the electron gap is nonzero across the transition. We argue below that this intermediate regime, where the electron gap remains nonzero on the disordered side is relevant to cuprates.

In the intermediate pairing regime, difficulties arise when we try to understand the electron spectrum when the superconductivity is disordered. Fermionic excitations in a superconducting state are Bogoliubov quasi-particles which are superpositions of electrons and holes. When the superconductivity is quantum disordered but close to the superconductor-insulator phase boundary, we expect by continuity that the insulator should have a band structure close to the Bogoliubov band. On the other hand, the charge conservation in an insulator prohibits the superposition of electrons and holes and seems to forbid a Bogoliubov-like band. The resolution of this puzzle lies in the small-gap bosonic pair, which exists when the system is close to but outside the superconducting phase. We discuss in Sec.~\ref{subsection: gapped sector} that part of the Bogoliubov band deforms into quasi-electron excitations, and the rest has to be understood as the two-particle continuum of a pair and a hole. As we further increase the insulating gap of the boson, the electron spectrum evolves adiabatically to that of a band insulator.

Now we go back to the fluctuating PDW in cuprates. It is experimentally observed that cuprate high-temperature superconductors have a very short coherence length, about 4 lattice spacing. It suggests the size of a pair is roughly comparable with the distance between neighboring pairs and the size of the MDW enlarged unit cell we consider; therefore the Coulomb repulsion between neighbouring pairs may drive the pairs into a Mott insulating phase. We propose the scenario that the anti-nodal electron gap is preserved when PDW is disordered, and the electron pairs form a Mott insulator in the MDW enlarged unit cell without further symmetry breaking. In Sec.~\ref{subsection: fluctuating PDW in cuprates} and Sec.~\ref{subsection: ARPES}, we apply the theory of a fluctuating fully gapped superconductor to describe the anti-nodal electron spectrum. 

Theoretically, the idea of a tight pair goes back to Anderson: roughly speaking, a hole in the $t-J$ model breaks a spin singlet nearby, two holes can avoid breaking two singlets by forming a pair, resulting in a pairing energy at a fraction of $J$. There has also been earlier discussions treating the anti-nodal pairs as bosonic preformed pairs that are coupled to the nodal electrons.~\cite{geshkenbein1997}.  

Unlike antinodal electrons, which are strongly paired under PDW, nodal electrons barely couple to the PDW because of momentum mismatch. (The PDW momentum $P$ is about twice the anti-nodal Fermi momentum; as seen from Fig ~\ref{Fig: Bogoliubov bands}a, it is considerably larger than the momentum that can be formed with a pair of electrons in the small Fermi pocket.) 
The nodal `arcs' are cut out and reconnected by the secondary CDW and remains largely unchanged by the PDW. Therefore while they are in principle  Bogoliubov bands, the gapless nodal bands can be viewed as electron bands weakly coupled to the PDW condensate. When the PDW disorders, the nodal bands go back to a pure electron band.

For the gapless bands coming from nodal electrons, the Bogoliubov-band paradox shows up in a different way. In the mean-field calculation (Fig.~\ref{Fig: Bogoliubov bands}), there are 2 gapless bands, hence 2 pockets, with identical shape, shifted by the PDW momentum, but the 4 `arcs' on the original Fermi surface can only form one closed pocket. From the perspective of total gapless degrees of freedom, the 2 pockets in the ordered PDW state is actually one pocket per spin, the same as we expect for the Harrison-Sebastian pocket. This is because the Nambu spinor representation $(c_{k\uparrow}, c_{P-k\downarrow}^{\dagger})^\text{T}$ already includes both spins, and puts down spin at shifted momenta. However, in the PDW-ordered state, due to the small but nonzero mixing of $c_{k\uparrow}$ and $c_{P-k\downarrow}^{\dagger}$, the gapless fermions acquire a nonzero spectral weight at PDW-shifted momenta, which should be absent in the PDW-disordered ground state. As we disorder the PDW, we need to explain how this extra spectral weight disappears. The answer is also rooted in the interplay between the bosonic pair and the electron, which we discuss in Sec.~\ref{subsection: gapless sector}.

In summary, by disordering the PDW, we arrive at a metallic state with a small electron pocket in the B.Z. folded by CDW and MDW. The extra charge density is carried by paired electrons which form a Mott insulator in the enlarged unit cell. The antinodal pairing gap is maintained. The state we are describing is adiabatically connected to a conventional small-pocket Fermi liquid with a large insulating gap of antinodal electrons. 

The reader may reasonably worry about the abrupt nodal-anti-nodal partition, for there is no sharp distinction between nodal and anti-nodal electrons on the original Fermi surface. Furthermore, for the above construction to work we need to partition the charge density, so that the bosonic pair is at commensurate density to form a Mott insulator, and the gapless pocket satisfies Luttinger's theorem. But the nodal electron pocket we start with is given by a mean-field PDW, which is a pairing state and does not satisfy Luttinger's theorem automatically. 

Our justification of this partition is twofold. First, the CDW descending from PDW cut the original Fermi surface into separate bands, so there is a natural distinction between nodal and anti-nodal electrons; second, the partition of density between the gapless fermion and the boson is a property of the energetics of the manybody ground state, which the mean-field PDW fails to address. Here we can only argue that such a partition is locally stable. Let us imagine that at some density, the gapless Fermi pocket  satisfies Luttinger's theorem in the reduced B.Z., consequently, the boson has integer filling consistent with the requirement of a Mott insulator. At low energies, the boson sector and the fermion sector effectively decouple. As we dope the system away from that density, it is energetically favorable for the extra electrons/holes to enter the gapless sector to avoid paying the Mott gap. Thus, the boson-fermion phase we considered is stable in a range of doping. Whether underdoped cuprates choose to partition its density this way, however, is an energetic question that can be tested only experimentally.

Next we check whether the available expereimental data are consistent with Luttinger's theorem. 
Although STM reports commensurate CDW of period 4 in a range of underdoped $\text{Bi}_2\text{Sr}_2\text{CaCu}_2\text{O}_{8+x}$ (Bi2212), resonant x-ray scattering and non-resonant hard x-ray diffraction report an incommensurate CDW in YBCO, with period smoothly passing through 3, and in $\text{HgBa}_2\text{CuO}_{4+\delta}$ (Hg1201), with period smoothly passing through 4.\cite{PhysRevB.96.134510} Whether a specific cuprate has incommensurate or commensurate CDW may depend on details like the strength of lattice-pinning, but the existence of CDW seems to be universal. Since Luttinger's theorem is a well defined concept only for commensurate superlattices, we restrict ourselves to commensurate CDW and PDW here. The incommensurate case will be viewed as comprising of commensurate domains. 

To compare with experiments, we identify the CDW momentum measured experimentally as twice the PDW momentum, and we check whether the pocket size measured from quantum oscillation obeys Luttinger's theorem at the specific doping when the CDW is commensurate. This kind of data is available only for the YBCO and Hg1201 systems, and within error bar, both YBCO and Hg1201 pass the test. According to Ref.~\onlinecite{PhysRevB.96.134510}, in YBCO, the CDW has momentum about $0.33*2\pi$ at 8\% doping, where the electron pocket is about  1.5\% of the original B.Z.,  accommodating 3\% of the electron density. The rest of the density, 0.92-0.03 = 0.89 per unit cell, is consistent with 16/18 = 0.89, ie 8 charge $2e$ bosons per MDW unit cell (which is 18 times the original unit cell).
\footnote{Equivalently, we can count the charges relative to half-filling, and say there is a pair of holes per MDW unit cell. These two countings are equivalent because the area of the MDW unit cell is an even multiple of the area of the original unit cell.}
In Hg1201, the experimental data is limited and we follow ref.~\onlinecite{PhysRevB.96.134510} to use their numbers based on the use of a parametrized band structure which they found to be in excellent agreement with the data.  The CDW has momentum about $0.25*2\pi$ at 12\% doping, where the electron pocket is about 4\% of the original unit cell, the rest of the density, 0.88 - 0.08 = 0.80 per unit cell, is consistent with 26/32 = 0.81, ie 13 bosons per MDW unit cell (which is 32 times of the original unit cell). 

The doping at which the CDW is commensurate can be determined experimentally with an error bar of roughly $1\%$, which is inherited from the error bar of the CDW momentum~\cite{PhysRevB.96.134510}. This uncertainty gives an uncertainty of the expected Fermi surface area, which is about $10\% \sim 15\%$ of the folded B.Z. We note that the test of Luttinger theorem is most sensitive to the doping density at a given commensurate doping, and the pocket size is only a small correction.  Thus Luttinger's theorem poses a highly nontrivial test to candidate theories as long as the doping density at a commensurate CDW momentum is  known with reasonable accuracy.

To further illustrate the nontriviality of the Luttinger theorem test, we note that the choice of the MDW unit cell is crucial. Since only  CDWs at 2P have been observed, one might be tempted to choose  $2P_x$ by $2P_y$ as the reduced BZ instead. In this case the real space unit cell is half the size of the MDW unit cell and we will have 6.5 bosons per unit cell for the Hg1201 case. This violates the integer density condition for the bosonic Mott insulator. In other words, if we form a bosonic Mott insulator in the CDW superlattice,  Luttinger's theorem will be strongly violated. 


In the next section, we follow the logic presented above to analyze the fluctuating PDW state in detail. We present mean-field PDW bands with different choices of order parameters. We construct simplified models to show how fermion spectral functions change as the PDW disorders, and to discuss how the bosons eat up the density of the fermions to form a Mott insulator. We then go back to the fluctuating PDW in cuprates and discuss experimental implications with insights from simplified models.

\section{Constructing the fluctuating PDW ground state}\label{Sec: Constructing the fluctuating PDW ground state}

In this section, we present the mean-field PDW bands, and address the questions of disordering the PDW step by step. We divide this section into five parts. Sec.~\ref{subsection: Mean-field PDW bands in cuprates} discusses the band structure of mean-field PDW and the symmetry of its descendant orders. Sec.~\ref{subsection: gapped sector} is on disordering an s-wave superconductor. Despite differences in pairing momentum and form factor, the physics of the electron gap and the interplay between electrons and pairs is essentially the same as the gapped sector of the fluctuating PDW. We focus on the electron spectral function in the disordered phase. Sec.~\ref{subsection: 1D numerics} presents numerical study of a 1D model, verifying the spectral features postulated in Sec.~\ref{subsection: gapped sector}, and illustrating how the boson adjust its density to form a Mott insulator. Sec.~\ref{subsection: gapless sector} discusses gapless PDW bands. Sec.~\ref{subsection: fluctuating PDW in cuprates} synthesis understandings of simple situations to address the fluctuating PDW in cuprates.

\subsection{Mean-field PDW bands in cuprates}\label{subsection: Mean-field PDW bands in cuprates}

A PDW condensate is a bath of charge 2e bosons carrying specific nonzero momenta. It mixes an electron with a hole, like regular superconductivity, but only at shifted momenta. To illustrate the PDW we consider in cuprates, we first sketch the band structure along the cut $k_y = \pi$, considering the effects of x-directional PDW and y-directional PDW separately. 

\begin{figure}[t]
\begin{center}
\includegraphics[width=0.95\linewidth]{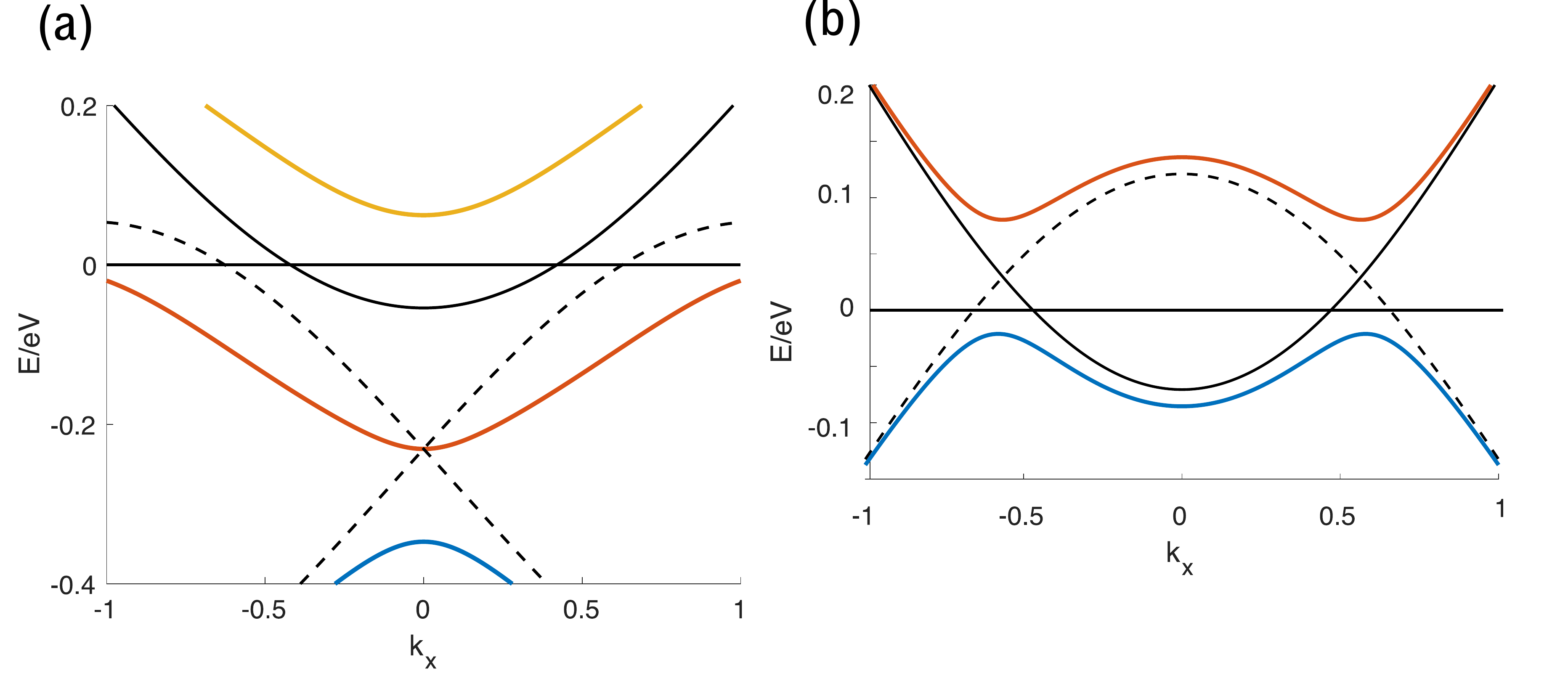}
\caption{Effects of x-directional PDW (a) and y-directional PDW (b) along the line $k_y = \pi$. The original electron band ($\e_k$) is shown as solid black line. PDW reflected bands ($-\e_{\pm P\hat{x}-k}$ and $-\e_{\pm P\hat{y}-k}$) are shown as dotted black lines. The hybridized Bogoliubov bands are shown in colors.}
\label{Fig: PDW 3-band illustration}
\end{center}
\end{figure}

Fig.~\ref{Fig: PDW 3-band illustration}(a) illustrates effects of x-directional PDW. We plot the energy of $c_{\vec{k}}$ (the original electron) as the solid black line, and energy of $c^{\dagger}_{\pm P\hat{x} - \vec{k}}$ as dashed black lines. PDW hybridizes these three bands into the red and blue bands below the Fermi energy, and the yellow band above the Fermi energy. Fig.~\ref{Fig: PDW 3-band illustration}(b) illustrates the mixing between $c_k$ and $c^{\dagger}_{\pm P\hat{y} -k}$ under y-directional PDW. In this case $c^{\dagger}_{P\hat{y} -k}$ and $c^{\dagger}_{-P\hat{y} -k}$ happen to be degenerate, and the electron band effectively couples to only their equal-weight superposition. Hybridization of the electron band and this superposition gives the red band and the blue band.
\footnote{The asymmetric superposition of $c^{\dagger}_{\pm P\hat{y}-k}$ does not couple to the electron; therefore appears to stay gapless. But this is an artifact of the 3-band approximation. For example, the coupling between this band and $c_{\pm 2P\hat{x}+k}$ can gap it} 
For bidirectional PDW, PDW in x-direction and PDW in y-direction together open a gap at antinodes, if the PDW amplitude is big enough. Which one dominates depends on details of the band structure, and the pairing momentum.

Different from what is reported in Ref.~\cite{PhysRevX.4.031017} (where the effect of the y-direction PDW was not considered), we find that y-directional PDW generically contributes more to the spectral gap at or near $k_y=\pi$. This feature can also be seen in the recent work of Tu and Lee. ~\cite{tu2019} In this scenario, as we gradually increase the PDW amplitude, the Fermi surface is gradually pushed towards larger absolute value of $k_x$ before the gap opens (Fig.~\ref{Fig: PDW 3-band illustration}(b)), while if the x-directional PDW dominates, we would see the Fermi surface pushed towards smaller $k_x$ and disappear at zero momentum (Fig.~\ref{Fig: PDW 3-band illustration}(a)).
In either case, as we move from $ky = \pi$ to $k_y = \pi/2$, at some point, PDW stops to provide a full gap. Because of momentum mismatch, PDW barely do anything to nodal electrons. For more details, see Ref.~\cite{PhysRevX.4.031017,PhysRevB.97.174511,PhysRevB.77.174502}. We remark that the addition of the y-direction PDW contribution shown in Fig.~\ref{Fig: PDW 3-band illustration}(b) has the desirable feature that the gap opens up for smaller pairing amplitude compared with the contribution from x-direction PDW alone.

\begin{figure*}[htb]
\begin{center}
\includegraphics[width=0.55\linewidth]{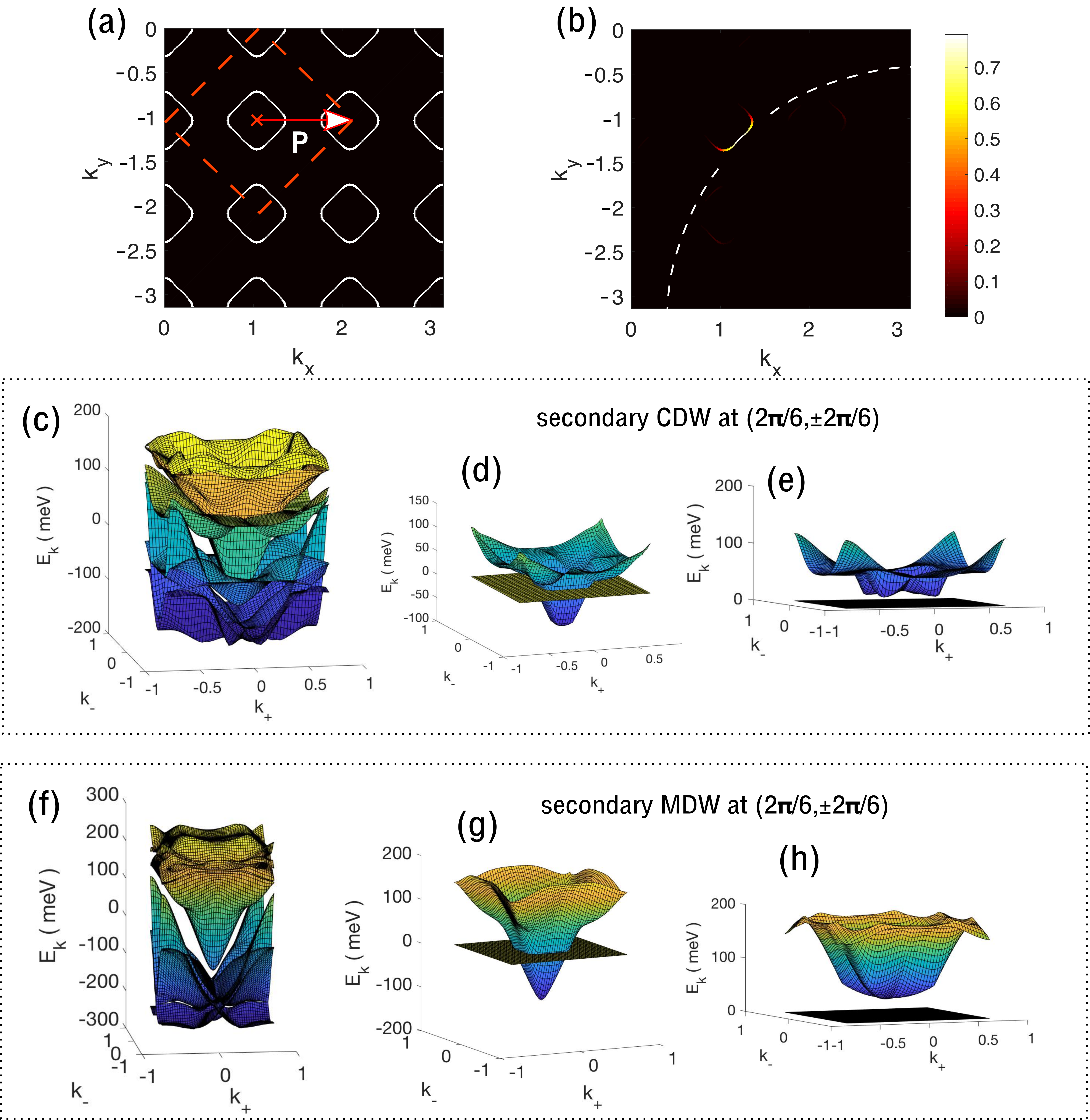}
\caption{Bogoliubov bands of ordered PDW. We use the mean-field Hamiltonian in Eq.~\ref{Eq: PDW mean field}, with hopping parameters $t = 154\text{meV}, t_p = -24\text{meV}, t_{pp} = 25\text{meV}, t_{ppp} = -5\text{meV}$ (see Eq.~\ref{Eq: tightbindingenergy}), chemical potential $\mu = -126\text{meV}$, PDW momentum $2\pi/6$, and PDW order parameter $|\D_P| = 40\text{meV}$. The original B.Z. is reduced to the small B.Z. spanned by $(\pi/3,\pm\pi/3)$. There are 36 bands coming from 18 electron bands and 18 hole bands in the reduced B.Z. Fig. (a): The gapless Fermi pocket. We plot the right-lower quadrants of the original B.Z., The red dashed line represents the reduced B.Z. The arrow represents PDW momentum ($P\hat{x}$ and $P\hat{y}$ are identical in the reduced B.Z.). Fig. (b): The spectral weight of zero-energy fermions in the original B.Z.. The white dashed line illustrates the large Fermi surface. Note that the new Fermi surface are mainly composed by the nodal portion of the original Fermi surface; its shape is barely changed by the PDW. Fig. (c): Bogoliubov bands close to Fermi energy. The PDW amplitudes are $\D_{P\hat{x}} = \D_{-P\hat{x}} = \D_{P\hat{y}} = \D_{-P\hat{y}}= 40\text{meV}$. This choice of phase produces CDW order at $(P,\pm P)$. $k_+$ and $k_-$ run between $\pm\pi/3\sqrt{2}$ along the diagonals. Bogoliubov bands appear in pairs: Each pair of bands have identical shape, they are related by a flip in energy (similar to the BCS bands) and a further shift by the PDW momentum. Fig. (d): the gapless band in Fig. (c). The horizontal plane represents the Fermi energy. Fig. (e): the first gapped band in Fig. (c). Fig. (f/g/h), the same as Fig. (c/d/e), except for $\D_{P\hat{x}} = \D_{-P\hat{x}} = \D_{P\hat{y}} = 40\text{meV}, \D_{-P\hat{y}}= -40\text{meV}$. This produces a magnetization density wave (MDW) state which orders at $(P,\pm P)$ and breaks time-reversal symmetry.}
\label{Fig: Bogoliubov bands}
\end{center}
\end{figure*}

In the analysis presented above, we have ignored higher order effects of PDW. For example, $c^{\dagger}_{P\hat{x} - k}$ also mixes with $c_{k - 2P\hat{x}}$. In general, we should consider the mixing between all of $c_{k + mP\hat{x} + nP\hat{y}}$ ($m+n$ even) and $c^{\dagger} _{-k + m'P\hat{x} + n'P\hat{y}}$ ($m'+n'$ odd). In this paper, we focus on the commensurate case with $P = 2\pi/6$, which is close to half of the CDW momentum in YBCO. The reduced B.Z. of non-superconducting density waves is spanned by $P\hat{x} \pm P\hat{y}$, with an area equal to 1/18 of the original B.Z. (red dashed square in Fig.~\ref{Fig: Bogoliubov bands}(a)). The 4 PDW momenta are all $(\pi,\pi)$ in the reduced B.Z.. The Hamiltonian we consider is 

\bea
H &=& \sum_{\vec{k},\s} \e_{\vec{k}} c^{\dagger}_{\vec{k},\s}c_{\vec{k},\s}\nonumber\\ 
&+& \sum_{\vec{k}} \D_{P\hat{x}}(\vec{k}) c_{\vec{k},\uparrow}c_{-\vec{k} + P\hat{x},\downarrow} + \D_{-P\hat{x}}(\vec{k}) c_{\vec{k},\uparrow}c_{-\vec{k} - P\hat{x},\downarrow} \nonumber\\
&+& \sum_{\vec{k}} \D_{P\hat{y}}(\vec{k}) c_{\vec{k},\uparrow}c_{-\vec{k} + P\hat{y},\downarrow} + \D_{-P\hat{y}}(\vec{k}) c_{\vec{k},\uparrow}c_{-\vec{k} -P\hat{y},\downarrow}\nonumber\\
&+& h.c.,
\label{Eq: PDW mean field}
\eea
where $\vec{k}$ runs in the original B.Z., and $\e_{\vec{k}}$ is the tight-binding dispersion: 

\bea
\label{Eq: tightbindingenergy}
\epsilon_k = &-&2t (\cos(k_x)+\cos(k_y)) - 4t_p\cos(k_x)\cos(k_y) \nonumber\\ &-&2t_{pp}(\cos(2k_x) + \cos(2k_y)) - \mu \nonumber\\ &-&4t_{ppp}(\cos(2k_x)\cos(k_y)+\cos(2k_y)\cos(k_x)).
\eea
For the choices of $t$, $t_p$, $t_{pp}$, $t_{ppp}$ and $\mu$, see the description of Fig.~\ref{Fig: Bogoliubov bands}. We choose a locally d-wave form factor for the PDW:

\bea
\label{Eq: PDW form factor}\D_{\vec{P}}(\vec{k}) = 2\D_{\vec{P}}[\cos(k_x-P_x/2) - \cos(k_y-P_y/2)]
\eea

As a general feature of the Nambu spinor representation, Bogoliubov bands of PDW shows up in pairs; each band has a partner that is flipped in energy and shifted by the PDW momentum.\footnote{For incommensurate PDW, we usually make an cutoff of higher order mixing which breaks this formal particle-hole symmetry (as shown in Fig.~\ref{Fig: PDW 3-band illustration}).} Of the 18 pairs of bands (coming from 18 electron bands and 18 hole bands), only 1 pair is gapless, giving 2 identical gapless Bogoliubov pockets in the reduced B.Z., shown in Fig.~\ref{Fig: Bogoliubov bands}(a).
\footnote{ All other bands are gapped out by the PDW as long as the PDW has a large amplitude and is bi-directional. See the description under Fig.~\ref{Fig: Bogoliubov bands} for details. Alternatively, we can reduce the PDW gap but explicitly add CDWs at momentum $2P$ to achieve similar results. On the other hand, bi-directional PDW is crucial in order to have only one pair of gapless bands. For a previous study of the band structure of unidirectional PDW with composite orders, see Ref.~\cite{PhysRevB.77.174502}. } 
However, the 2 pockets represent the same excitations. Counting the degrees of freedom, there is only one gapless pocket per spin. The reason is that the Nambu spinor representation shifts the down spin electrons by the PDW momentum, causing a superficial doubling. Physically, there are 2 pockets related by $(\pi,\pi)$ in the reduced B.Z. because momenta is conserved only up to $(\pi,\pi)$ when PDW is ordered. We shall see in Sec.~\ref{subsection: gapless sector} that after disordering the PDW, only the pocket at the center of the B.Z. left. The other pocket becomes a broad 2-particle continuum with a small gap.

Fig.~\ref{Fig: Bogoliubov bands}(b) shows the spectral weight of zero-energy electrons in the original B.Z.. We can see that gapless excitations come solely from nodal electrons along the original Fermi surface; anti-nodal electrons are all gapped. The CDW generated by the PDW connects the gapless arcs to form a closed pocket. Note that the effect of zone-folding in electron spectral function is visible only at the tips of the nodal arc, due to the fact that the CDW amplitude is much smaller than the hopping. On the contrary, if we were to gap out anti-nodal electrons by only CDW, we would need a CDW amplitude comparable to the hopping, resulting in an unrealistically large mixing between $c_k$ and $c_{k+2P}$.

By the approximate $C_4$ symmetry of the $\text{CuO}_2$ plane, we assume the 4 PDW order parameters in Eq.~\ref{Eq: PDW mean field} have about the same amplitude. However, different choices of the 4 phases give different ground-state energies and symmetries~\cite{agterberg2008dislocations}. Of the 4 phases, we can use the $U(1)$-charge symmetry to fix one. In the limit that PDW wavelength is much bigger than the lattice spacing, we can use continuous translation in x and y direction to fix two more phases. In this case, the only nontrivial phase is $e^{i\theta}\equiv \D_{P\hat{x}}\cdot\D_{-P\hat{x}}/(\D_{P\hat{y}}\cdot\D_{-P\hat{y}})$. Time reversal symmetry requires it to be 1. Any other choice breaks time reversal (spontaneously). Fig.~\ref{Fig: Bogoliubov bands}(c) and Fig.~\ref{Fig: Bogoliubov bands}(f) shows the 8 bands close to Fermi energy for $\theta = 0$ and $\theta = \pi$ correspondingly. The time-reversal invariant case ($\theta = 0$) has a CDW at momentum $(2\pi/6, \pm 2\pi/6)$ (App.~\ref{Appendix: Symmetry of the fluctuating PDW state}), which is apparently excluded by current experiments. The time-reversal breaking case ($\theta = \pi$) has a more stable band structure with a larger gap for the gapped bands (Fig.~\ref{Fig: Bogoliubov bands}(h)). In this case, the secondary order generated by PDW at momentum $(2\pi/6, \pm 2\pi/6)$ is purely current modulation without charge modulation. This orbital magnetization density wave (MDW) may also break the mirror symmetry along the diagonal. In each case, the specific band gap depends on band structure and PDW order parameters, but the nodal pocket and the shape of bands are more robust. See Ref.~\cite{agterberg2008dislocations} and App.~\ref{Appendix: Symmetry of the fluctuating PDW state} for details on the symmetry of the commensurate and incommensurate PDW.

\subsection{Fluctuating s-wave superconductor}\label{subsection: gapped sector}

Disordering the mean-field PDW ansatz with 36 bands is not an easy task. In this sub-section, we discuss a simplified model for the gapped sector of the fluctuating PDW: fluctuating s-wave superconductor. The intriguing feature of the fluctuating PDW state proposed in Sec.~\ref{Sec: big picture} is that although the anti-nodal gap comes mainly from PDW instead of the secondary MDW or CDW, PDW leaves no sign of further symmetry breaking since it is disordered. The paired electrons form an insulator instead of a superconductor. To understand this pairing induced insulator, we first discuss the  disordering of an s-wave superconductor with 2 electrons per unit cell, to see how an insulator emerges that preserves the lattice symmetry. Despite differences in the pairing momentum and local form factors, the interplay between pairs and fermions in the simplified model is essentially the same as in the fluctuating PDW. 

As introduced in Sec.~\ref{Sec: big picture}, there are several different regimes of the disordered superconductor as we vary the strength of the pairing. 

In the strong pairing regime (BEC limit), the binding energy of the electron pair is much larger than the Fermi energy. The superconductor with 2 electrons per unit cell (in average) is essentially the superfluid phase with 1 boson per unit cell (in average). Increasing the repulsion of the pairs, we can disorder the superconductor to get a bosonic Mott insulator, which is adiabatically connected to the atomic insulator with one pair per unit cell. The effective theory near the superconductor-insulator phase transition is the 3D XY model. It is clear that only the bosonic gap closes at the transition; the electron gap, which is essentially the binding energy of the pair remains large across the transition.

In the weak pairing regime, the superconducting phase is well-described by the BCS theory; and the size of pairs is much larger than the lattice spacing. Therefore, it is not clear whether the Mott insulator of pairs can be energetically favorable when we disorder the superconductor. The single-electron gap may not persist to the disordered side.

We are interested in the intermediate pairing regime, where the pairing amplitude is comparable or smaller than the Fermi energy but not too small. We expect by continuity from the BEC limit that the transition from the superconductor to a bosonic Mott insulator still exists, and the universality class is unchanged. However, there seems to be a paradox related to the fermion spectrum. In this intermediate regime, the fermionic excitations in the superconducting phase are Bogoliubov quasi-particles which roughly follow the BCS bands. In the insulating phase but close to the transition, we expect by continuity that the band structure of the insulator should be similar to the Bogoliubov bands. This expectation seems to contradict the charge conservation, which forbids the mixing between electron bands and hole bands.

In the rest of this subsection, we solve this puzzle of Bogoliubov bands and build intuition on the pairing induced insulator in the intermediate pairing regime. For concreteness, we imagine a metal with 2 bands per spin, each half-filled, to give 2 electrons per unit cell. Under s-wave pairing, the Fermi surface is fully gapped. We then disorder the bosonic pair at low energy while maintaining the pairing to get the bosonic Mott insulator. On the insulating side, close to the transition (where the boson gap closes), we are in the limit that the gap for charge 2e bosonic excitations (which we call $\D_b$) is much smaller than the gap for charge e fermionic excitations (which we call $\D_f$), and they are both smaller than the Fermi energy:
\bea 
\D_b \ll \D_f< E_F.
\eea

For energy scales much smaller than $\D_f$, we cannot excite any fermion; the system is effectively a bosonic system, and all charges are carried by bosons in the low-energy effective description. We then tune the boson interaction at this length scale to drive it to a Mott insulator with a small gap $\D_b$. Note that this procedure can be done most effectively when the range of interaction is comparable to the size of the boson. More physically, each bosonic pair we consider in cuprates spans around 4 lattice spacing, comparable to the MDW enlarged unit cell, but still has considerable overlap with neighboring pairs. We are naturally in the limit where a Mott gap starts to be possible, and it has to be small if there is any. 

Note that we cannot get the desired insulator by treating pairing perturbatively. If we start from a Fermi liquid, and calculate the self energy correction by coupling to a small-gap charge-2e boson, we can at most get a Fermi surface with reduced spectral weight~\cite{PhysRevB.79.245116}. The reason is simply that to connect the unoccupied electrons well-above the Fermi level, and the occupied electrons well-below the Fermi level, the real part of the corrected self energy must change sign by going through zero, hence giving a Fermi surface.
\footnote{In principal, the self energy may also diverge, as the BCS self energy, but it is not possible when the boson is gapped. In fact, such a divergence signals the breakdown of the perturbation.}

In fact, the key feature that makes this insulator easy to understand is precisely that the charge 2e boson gap $\D_b$ is much smaller than the fermion gap $\D_f$. We may compare this feature with a superconductor, where $\D_b=0$ (ignore Coulomb interaction), or with a free-electron insulator, where the lowest bosonic excitation is just the 2-electron excitation at the band minimum, hence $\D_b = 2\D_f$. Interestingly, this pairing-induced insulator is adiabatically connected to a trivial band insulator, but energetically closer to a superconductor.

When the pair excitation gap is much smaller than the single fermion gap, band theory cannot give a satisfactory description. As an effective field theory, we use a complex boson field $\phi$ to describe low energy pair excitations, and a fermion operator to create a gapped unpaired electron. At low energy, the bosonic action should be quadratic in time since it has integer filling per unit cell~\cite{sachdev_2011}.

\bea
\label{Eq:bosonLagrangian}
\mathcal{L}_b &=& \frac12|\partial_t\phi|^2 - \frac12 v_b^2|\nabla\phi|^2 - \frac12 \D_b^2|\phi|^2\\
\mathcal{H}_b &=& \sum_k E^b_k (a^\dagger_ka_k + b^{\dagger}_kb_k),
\eea
where we use canonical quantization to write $\phi(p) = \frac{1}{\sqrt{E^b_p}}(a_p + b_{-p}^{\dagger})$, and $E^b_k = \sqrt{\D_b^2 + v_b^2k^2}$ for small $k$. $\phi_p$ carries charge 2e; $b_p$ and $a_p$ are the annihilation operators of the bosonic pair and the vacancy of pair.

\begin{figure}[htb]
\begin{center}
\includegraphics[width=0.6\linewidth]{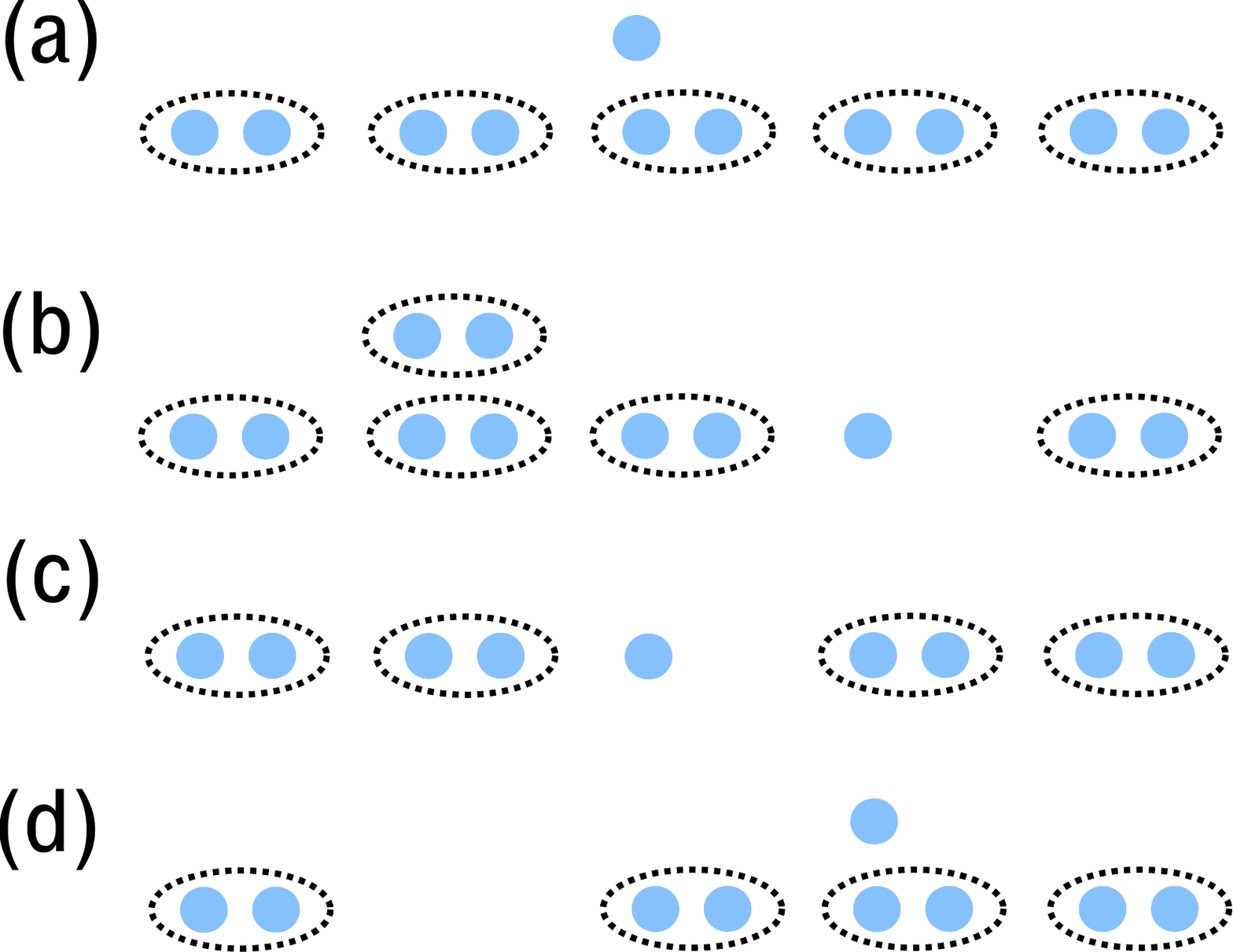}
\caption{Fig. (a) and (b): sketch of excitations created by adding an electron. Fig. (c) and (d), sketch of excitations created by removing an electron. In ARPES experiments, the incident photon may break a pair and create a hole (Fig. (c)); it may then decay into the continuum of an electron and a boson vacancy as illustrated in Fig. (d).}
\label{Fig: electron hole continuum}
\end{center}
\end{figure}

As illustrated in Fig.~\ref{Fig: electron hole continuum}, the basic excitations in this system are electrons, holes, pairs and vacancies of pairs. Contrary to our usual intuition, pairs and vacancies of pairs are well-defined quasi-particles in this insulator for they are the lowest charged excitations. For energy scales below $\Delta_f$, the bosonic theory in Eq.~\ref{Eq:bosonLagrangian} is the complete description of low energy excitations.

Since a fermion cannot decay into a boson, electron excitations and hole excitations can still be quasiparticles even though $\D_f$ is much larger than $\D_b$. However, the electron and hole spectra are strongly affected by the low-energy boson; therefore they are very different from the spectrum of a band insulator. As illustrated in Fig.~\ref{Fig: electron hole continuum}(a) and Fig.~\ref{Fig: electron hole continuum}(b), when we add an electron to the system, it may either be a single electron (Fig.~\ref{Fig: electron hole continuum}(a)), or split into a hole and a pair (Fig.~\ref{Fig: electron hole continuum}(b)). Since these two configurations have the same electric charge, an eigen-state of the charge e excitation is always a mixture of the two. In fact, the single electron in Fig.~\ref{Fig: electron hole continuum}(a) is just the special case of Fig.~\ref{Fig: electron hole continuum}(b), where the hole and the pair overlap. Thus, whether the addition of an electron creates a quasiparticle excitation depends on whether the hole and the pair in Fig.~\ref{Fig: electron hole continuum}(b) form a bound state. The physics for removing an electron is similar, as illustrated in Fig.~\ref{Fig: electron hole continuum}(c-d). This line of thinking is particularly useful in the current case, where the boson gap is small. Since the energy of the bosonic pair is small around zero momentum, if the electronic excitation has lower energy than the hole excitation at momentum $k$, the electronic excitation likely form a quasiparticle, but the hole excitation is no longer a quasiparticle: it decays into the two-particle continuum with an electron near momentum $k$ and a boson near momentum $0$.

\begin{figure*}[htb]
\begin{center}
\includegraphics[width=0.6\linewidth]{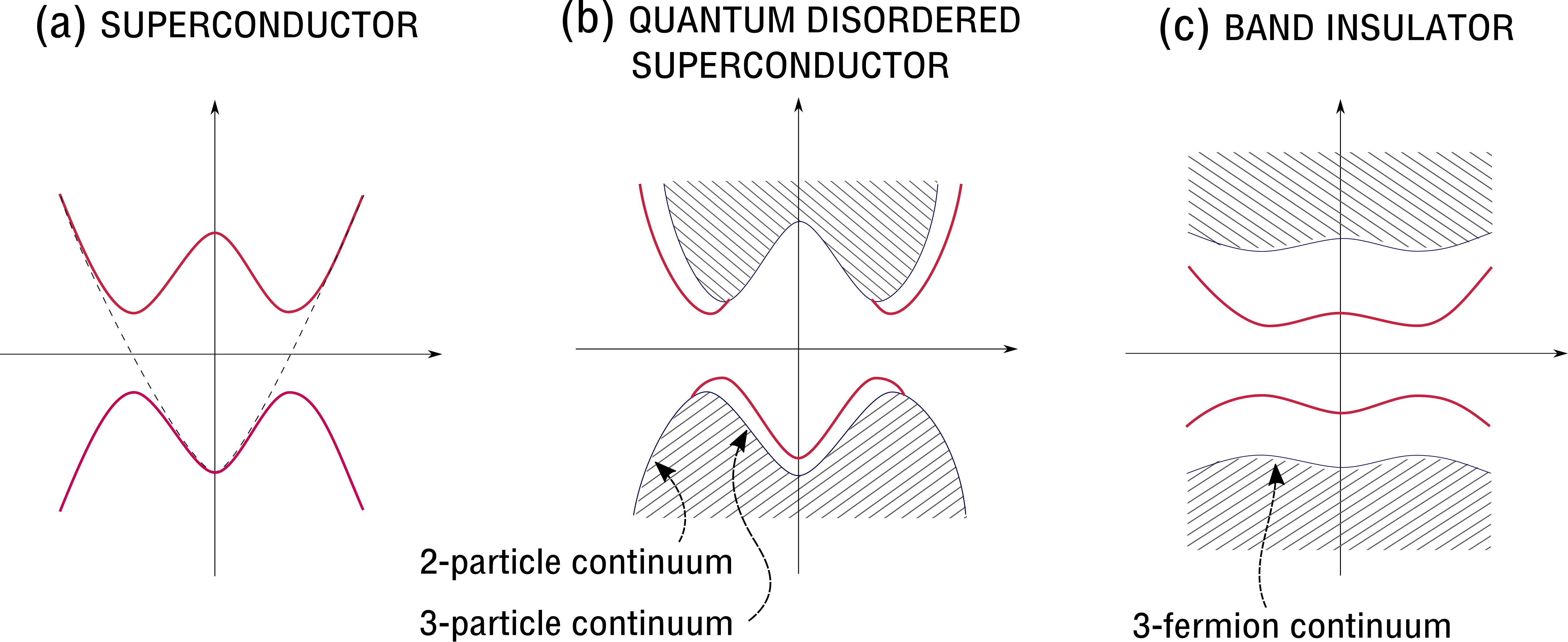}
\caption{Evolution of fermionic excitation from an s wave superconductor to an insulator. Fig. (a): The BCS band of an s wave superconductor (solid red line). The original electron band before pairing is shown as the dashed line. Fig. (b): Electron band (solid red line) and the boson-fermion continuum (shaded area) when the superconductor is quantum disordered but close to the transition point, $\D_b \ll \D_f < E_F$. The multi-particle continuum here plays a more important role than in usual insulator, because the bosonic pair has a small energy gap when it is close to condensing. The quasi-electron band and the quasi-hole band (solid red line), together with the k-dependent threshold of the 2-particle continuum (shaded) together resembles the BCS band. Fig. (c): electron and hole band in a usual band insulator (solid red line) and the 3-fermion continuum (shaded area). We can smoothly interpolate between Fig. (b) and Fig. (c): as we increase the boson gap, the boson-fermion continuum gradually separates from single-fermion excitations. Eventually, the electron band has little resemblance of the BCS band, the boson fades into the 2-fermion continuum, and the boson-fermion continuum becomes the 3-fermion continuum.}
\label{Fig: fluctuating SC band}
\end{center}
\end{figure*}

In order to understand the fermionic spectrum of the insulator in the limit $\D_b\ll \D_f < E_F$ , we first look at the BCS bands of the superconductor.

\bea
H_\text{f,BCS} = \sum_k (c^\dagger_{k\uparrow}, c_{-k,\downarrow})
\left( \begin{array}{cc}
\epsilon_k & \D_f \\
\D_f & -\epsilon_{-k}
\end{array} \right)
\left(\begin{array}{c}
c_{k,\uparrow}\\ c^\dagger_{-k,\downarrow}
\end{array} \right)
\eea

The fermionic excitations are Bogoliubov quasiparticles with energy
\bea
E^f_k = \sqrt{\e_k^2 + \D_f^2}
\eea

When the boson is barely disordered, we expect the fermionic spectrum to roughly follow the Bogoliubov bands but with two important changes: (1) excitations should now carry definite charges, (2) there may not be quasiparticle excitations at all momenta in this strongly interacting limit. No matter whether there is a quasiparticle or not at a specific momentum $k$, there is always an energy threshold for manybody states with charge $\pm e$ and momentum $k$. When there is a quasi-electron, there is a single state at the threshold instead of a continuum of states; in this case, we define the excitation energy of the quasi-electron to be $E^{e}_k$. Similarly, we define the excitation energy of the quasi-hole to be $E^{h}_k$, if it exists at momentum $k$. By definition, $E^{e}_k, E^{h}_k > 0$. To be consistent with conventions in free electron band theory, we plot $E^{e}_k$ and $-E^{h}_k$, to put charge e excitations in the upper-half plane, and charge -e excitations in the lower-half plane (Fig.~\ref{Fig: fluctuating SC band}). 

When the pairing is smaller than the band width, by continuity, we postulate Fig.~\ref{Fig: fluctuating SC band}(b) as the band structure of the insulator. For momenta  away from the band minimum and larger than the original Fermi momentum, we have the usual electron as a quasi-particle, with energy $E^{e}_k$ slightly distorted from the dispersion of the metal by pairing (Fig.~\ref{Fig: fluctuating SC band}(b), solid red curve in the upper plane). 
\footnote{It may decay into 3 fermions when $E^{e}_k>3\D_f$, but we ignore this usual decaying process for now.} 
There is no way to excite a hole at these \textit{unoccupied} momenta, but we can create an electron and remove a zero-momentum pair, hence a 2-particle continuum for hole excitations starting roughly from the energy $E^{e}_k + \D_b$.
\footnote{Here we assume the boson velocity is not too small, so the energy for bosonic excitation is small only near zero momentum.}
Similarly, for momenta smaller than the original Fermi momentum and away from the band minimum, we have quasi-holes with the energy $E^{h}_k$ (Fig.~\ref{Fig: fluctuating SC band}(b), solid red curve in the lower plane) and a 2-particle continuum for electron excitations starting roughly from $E^{h}_k + \D_b$. Near the band minimum (at the original Fermi surface), we should have at least one of the quasi-electron and quasi-hole, because the lowest fermionic excitation cannot decay into other particles. Since, the electron and hole dispersion are approximately symmetric near the band minimum, we should have a range where quasi-electron and quasi-hole coexist. 

As we follow the electron band from outside the Fermi surface to inside the Fermi surface (in Fig.~\ref{Fig: fluctuating SC band}(b)), the quasi-electron excitation starts to transition from a single electron depicted in Fig.~\ref{Fig: electron hole continuum}(a) to a bound state of hole and continuum depicted in Fig.~\ref{Fig: electron hole continuum}(b). After passing the band minimum, the excitation energy goes up, and the bound state become weaker, and finally the hole and pair no longer bind together, and the quasi-electron fades into the 2-particle continuum. The unbinding transition happens when $E_k^e = \text{min}_p\{E_p^h + E^b_{k-p}\}$. Deep in the Fermi sea, electron excitations do not make sense, and there is not even a resonance above the 2-particle continuum.

The quasi-particle band, together with the threshold of the 2-particle continuum resembles a BCS band. In addition, at energies $2\D_b$ above each quasi-particle excitation, we have a 3-particle continuum of one fermion and a particle-hole pair of bosons. Multi-particle continuum plays an important role in the insulator we discussed because of the small gap of the bosonic pair.

As we drive the insulator farther away from the critical point, the boson gap increases, and the fermion band gradually separates from the boson-fermion continuum. Eventually, the boson gap is so large that it fades into the 2-fermion continuum, and we arrive at a usual band insulator (Fig.~\ref{Fig: fluctuating SC band}(c)).

\begin{figure}[htb]
\begin{center}
\includegraphics[width=0.9\linewidth]{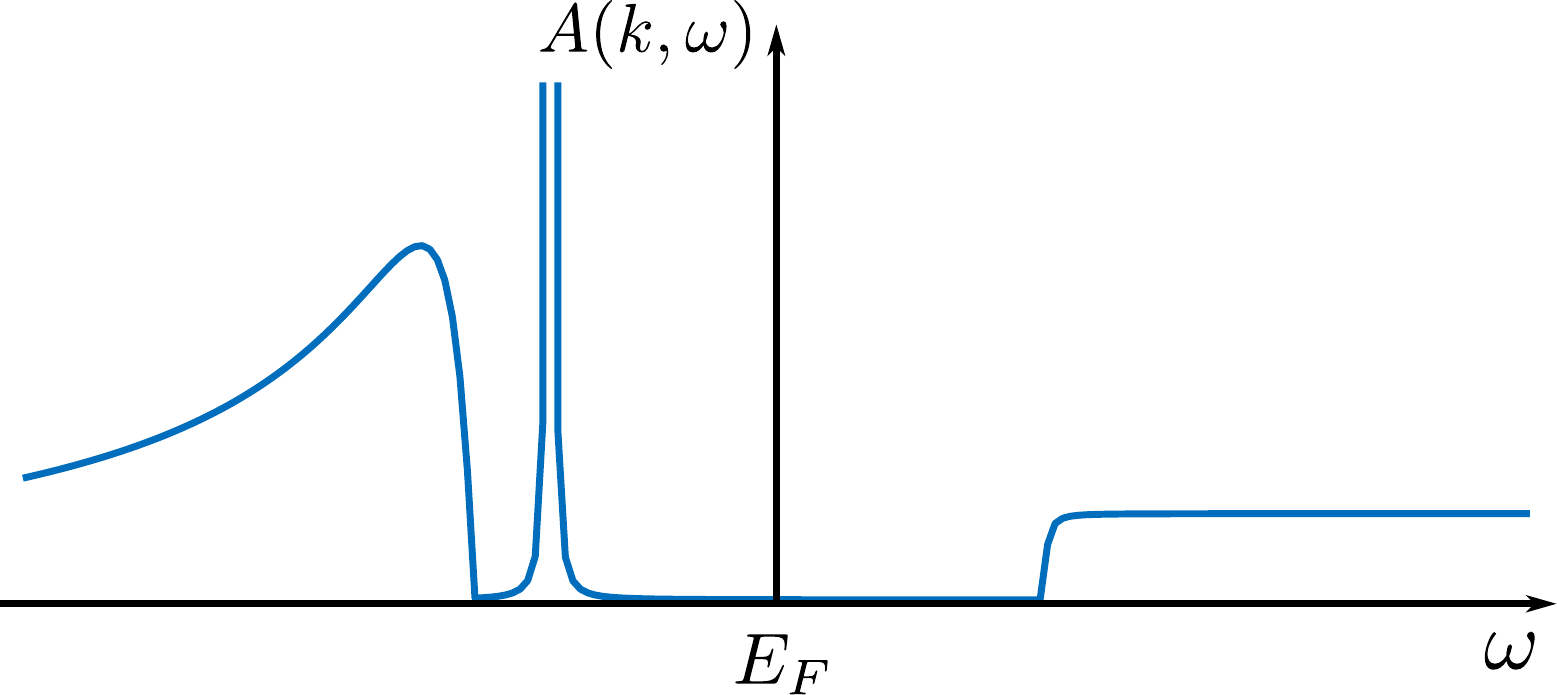}
\caption{Sketch of the electron spectral function at some $k<k_F$. Below $E_F$ there is a quasi-hole peak (delta function  in the ideal case, broadened here for the purpose of illustration.) and a 3-particle continuum. Above $E_F$ there is a 2-particle continuum. More detailed calculations in Sec.~\ref{subsection: ARPES} show that the 2-particle continuum onsets as a step function, while the 3-particle continuum decreases as $1/\omega$ for large frequencies.}
\label{Fig:sketchspectrum}
\end{center}
\end{figure}

To further illustrate the unconventional spectral features of this pairing-induced insulator, we sketch the spectral function for a fixed momentum $k<k_F$, where only quasi-hole exists. See Fig.~\ref{Fig:sketchspectrum}. We shall discuss the spectral features of the multi-particle continuum in more details in comparison with ARPES in Sec.~\ref{subsection: ARPES}.

We would like to comment that we present a non-perturbative understanding of fluctuating orders, a way to open a gap on Fermi surface without breaking any symmetry. Our discussion is general; whether the resulting state is energetically favorable or not depends on details. With special care of the charge and momentum carried by the fluctuating boson, similar arguments apply to other fluctuating orders, e.g. PDW, CDW and SDW, if the boson gap is much smaller than the fermion gap. The common feature is that quasiparticle peaks exist only in part of the B.Z., and it must be replaced by boson-fermion continuum in the rest of B.Z.. For fluctuating PDW, the boson has a small energy near a finite momentum $P$; electron at momentum $k$ and hole at momentum $k-P$ compete: if one of them has smaller energy, the other likely falls into the boson-fermion continuum.

\subsection{Pairing-induced insulator in 1D}\label{subsection: 1D numerics}

\begin{figure}[htb]
\begin{center}
\includegraphics[width=0.95\linewidth]{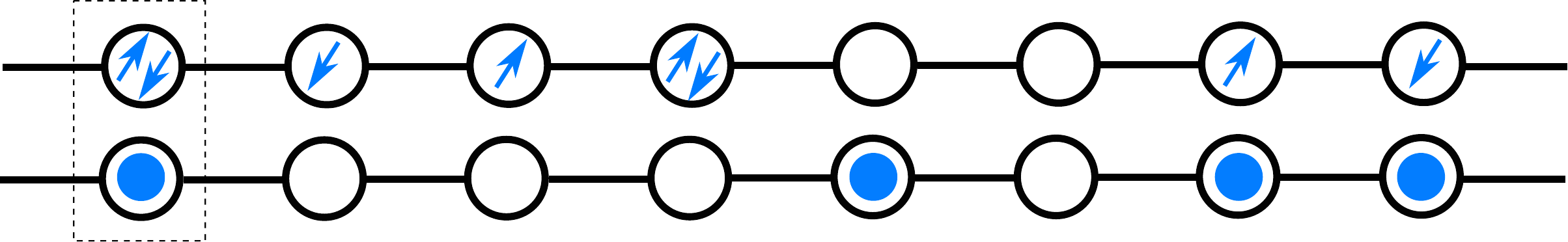}
\caption{1D boson-fermion model. Blue dots represent charge-2 hardcore bosons, blue arrows represent spin-up and spin-down fermions}
\label{Fig: 1D lattice}
\end{center}
\end{figure}

To test the idea of pairing induced insulator and its electron spectral function discussed in Sec.~\ref{subsection: gapped sector}, we design a simple 1D model, with charge-1, spin-1/2 fermion $c_{i\s}$ and charge-2, hardcore boson $b_{i}$. As illustrated in Fig.~\ref{Fig: 1D lattice}, each unit cell can have a spin-up fermion, a spin-down fermion, and a hardcore boson, independently. The Hilbert space for each unit cell is 8-dimensional. We choose the Hamiltonian to be:
\bea
H = &-&t_c\sum_{\<ij\>, \s}c^{\dagger}_{i\s}c_{j\s} -t_b\sum_{\<ij\>}b^{\dagger}_{i}b_{j}\nonumber\\
&+& \Delta\sum_{i} b_{i}^{\dagger}c_{i\uparrow}c_{i\downarrow} + h.c. + U\sum_{i} P^{0,4}_{i}
\label{Eq: 1D Hamiltonian}
\eea
where $P^{0,4}_i$ is the projector that is 1 if the $i$th unit cell contains total charge 0 or 4. This Hamiltonian conserves The total charge
\bea
Q = \sum_{i} 2 b^{\dagger}_{i}b_{i} + \sum_{i,\s}c^{\dagger}_{i\s}c_{i\s}.
\eea
There is an overall particle-hole symmetry that pins the total filling to charge-2 per unit cell. (Both the fermion and the hardcore boson are, on average, half-filled.) If $\<b_i\>\neq 0$, the $c_{i\s}$ fermion forms a proximity-induced 1D superconductor~\footnote{In a pure 1D system, we never have $\<b_i\>\neq 0$, but at best a power-law order.}.
What interests us is that even with this purely 1D model, with $b_i$ disordered, the pairing term still opens a fermion gap, but drives the system into an insulating state (for a range of $U$). To make connection with real materials, we can think of the boson as describing well-developed fermion pair of another band. We use this fermion-boson model instead of an all-fermion model, both for numerical convenience, and to illustrate how boson and fermion exchange density dynamically.

The physics of the pairing can be understood as follows. In the free theory, $\D = U = 0$, the left-moving and right-moving electron operator $c_{L,\sigma}$ and $c_{R,\sigma}$ have scaling dimension $1/2$. Without further interaction, the hardcore boson corresponds to a free fermion under Jordan-Wigner transformation, and $b^{\dagger}\sim e^{i\phi}$ has scaling dimension 1/4.
\footnote{We can determine the scaling dimension of the boson operator by bosonization. Write $b^{\dagger}\sim  e^{i\phi}$, and the corresponding left-moving and right-moving fermion after Jordan-Wigner transformation as $f_{R/L}^{\dagger} = e^{i(\phi \pm \theta)}$. As free fermion operators, $f_L $ and $f_R$ have scaling dimension 1/2, and $f_Lf_R\sim e^{2i\phi}$ has scaling dimension 1. Thus, $b^{\dagger}\sim e^{i\phi}$ has scaling dimension 1/4.}
Thus the pairing interaction $b^{\dagger}c_{\uparrow}c_{\downarrow}$ has scaling dimension $5/4$ and is relevant. The gapless fermion is unstable to pairing. The pairing renormalizes the bare boson operator $b$ into $\tilde{b} \sim u b+ v c_{\uparrow}c_{\downarrow}$. A single electron with no partner to form a pair fails to make the superposition with the boson, resulting in a pairing gap. Below this pairing gap, the model is effectively a model of the renormalized boson. The renormalized boson takes the density of both the bare boson and the fermion pairs below Fermi surface, becoming filling 1 per unit cell at low energies. Adding infinitesimal $\Delta$ immediately draw the system from the independent boson-fermion Luttinger liquids, to a one-component bosonic Luttinger liquid at low energy. Whether the bosonic Luttinger liquid is stable depends on the renormalized bosonic repulsion.

\begin{figure*}[htb]
\begin{center}
\includegraphics[width=0.7\linewidth]{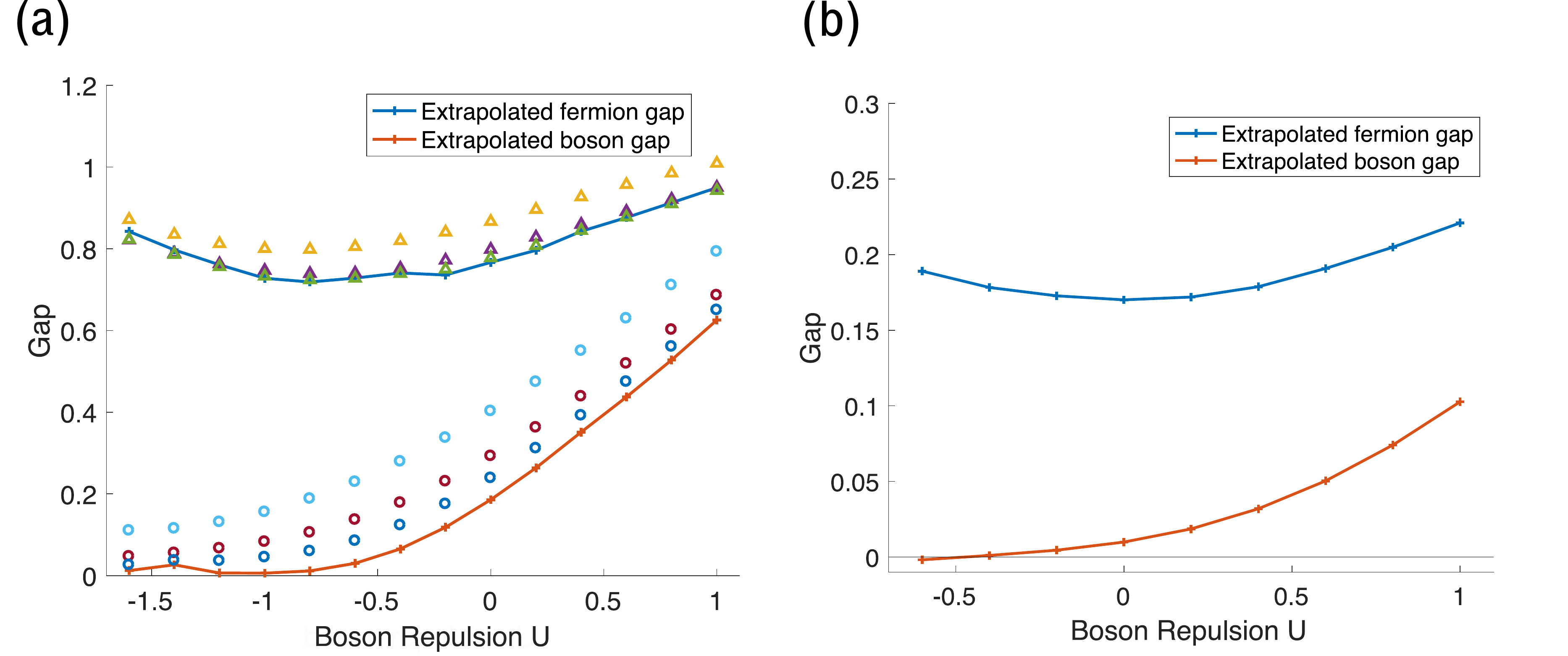}
\caption{(a) fermion gap (blue `+') and boson gap (red `+') of the 1D model, extrapolated from finite size DMRG calculation with system size $L=10,20,40$, $t_b = t_c =1,\D=1.3$, total filling: charge-2 per unit cell. Fermion gaps for $L =10$ (yellow triangle), $L=20$ (purple triangle), $L=40$ (green triangle), and boson gaps for $L = 10$ (light blue circle), $L = 20$ (dark red circle), $L=40$ (dark blue circle) are shown for reference. (b) The same as (a) except for small pairing $\D=0.5$. Finite-size extrapolation shown in Appendix \ref{Appendix: DMRG}. In both cases, the ground state go through a transition from a bosonic Luttinger liquid to a bosonic Mott insulator.  Fermion gap stays open across the transition.}
\label{Fig: 1D fermion boson gap}
\end{center}
\end{figure*}

\begin{figure}[htb]
\begin{center}
\includegraphics[width=0.8\linewidth]{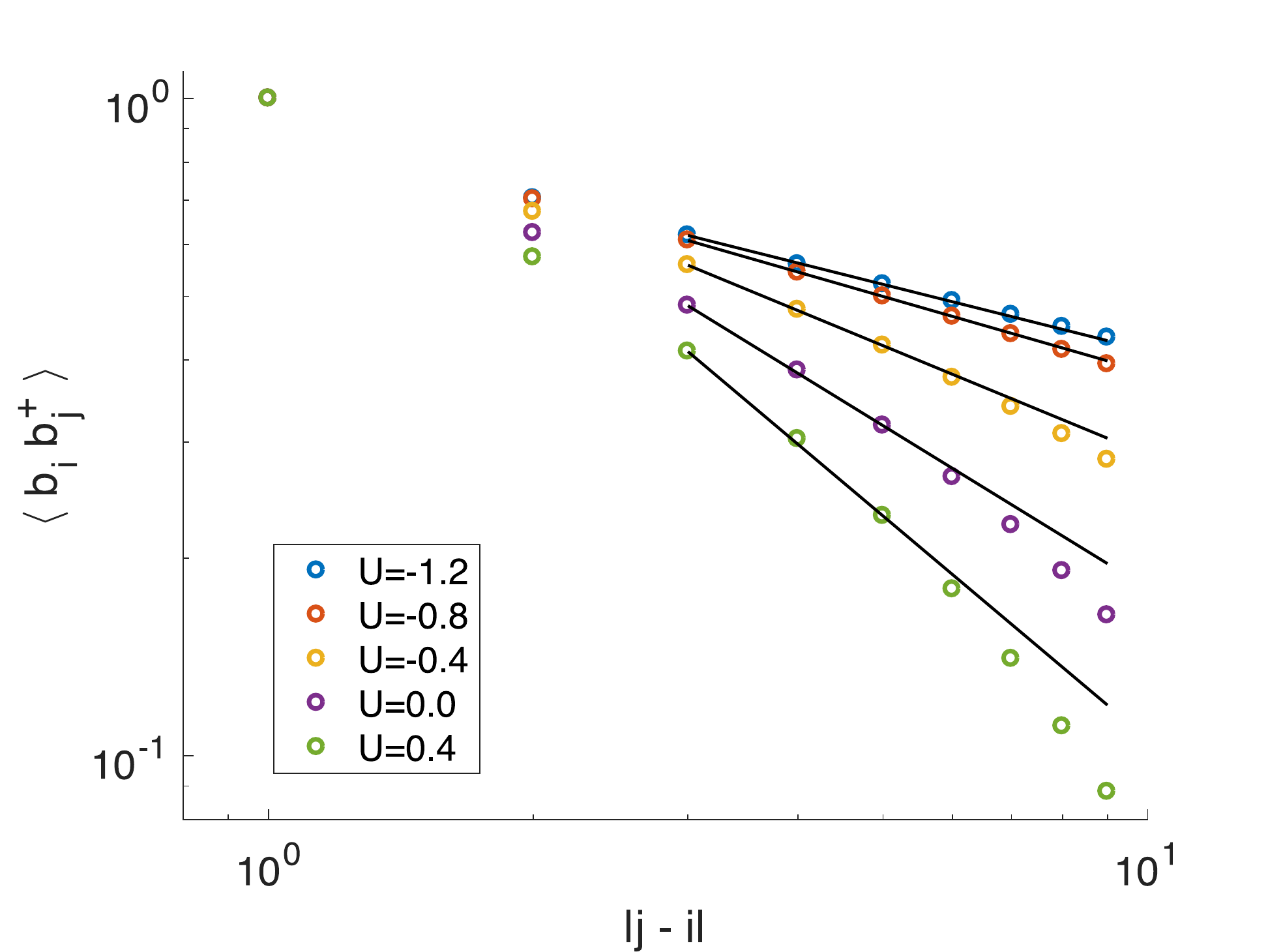}
\caption{Boson correlator $\< b_ib^{\dagger}_j\>$, in log-log scale. We use the ground state calculated by DMRG for $L=40$, fix $i=16$, and scan $j = 17,18,\dots, 25$. Black lines are guides to the eye. The correlator decays as power-law for $U=-1.2,-0.8$, but faster than power-law for $U=-0.4,0.0,0.4$, consistent with the gap calculated by DMRG.}
\label{Fig: 1D boson correlator}
\end{center}
\end{figure}

By tuning the bosonic Hubbard $U$, we can realize 3 different phases. For large repulsive $U$, we should have a bosonic Mott insulator in 1D, with charge 2 per unit cell. The state on each site is a superposition between the fermion pair and the bare boson.
(Since translation and particle-hole symmetry is maintained, the average occupation of the bare boson is 1/2 per site.)
For a range of attractive $U$, the renormalized boson forms a charge-2 Luttinger liquid. Single fermion is gapped, but the pair is gapless, realizing a Luther-Emery liquid. For large attractive $U$, we either have a CDW or phase separation. The charge on each site wants to deviate from 2, either smaller or larger. Note that no matter what $U$ is, single fermion is always gapped by the pairing. By design, the original boson itself has average filling $1/2$ and it is impossible to form a Mott insulator on its own. Seeing an insulator that preserves the translation symmetry implies that the boson has absorbed all the fermions to increase its effective filling to 1. We are interested in the transition between the Luther-Emery liquid and the Mott insulator, i.e., the emergence of the insulating phase with a small Mott gap.

We calculate the approximate ground state by DMRG for systems with length $L = 10,20,40$. We consider two cases, with large pairing ($t_b = t_c =1, \Delta = 1.3$) and relatively small pairing ($t_b = t_c =1, \Delta = 0.5$). In each case, we scan $U$ to drive the system from the bosonic Luttinger liquid to the pairing-induced insulator. For all parameters shown in Fig.~\ref{Fig: 1D fermion boson gap} and Fig.~\ref{Fig: 1D boson correlator}, we find that translation symmetry is preserved in the bulk. In the large pairing case (Fig.~\ref{Fig: 1D fermion boson gap}(a)), the extrapolated boson gap (red `+') is zero within the error bar for approximately $U\le -0.8$, and nonzero above that, indicating a continuous phase transition into an insulating ground state (see also the boson correlator in Fig.~\ref{Fig: 1D boson correlator}). On the other hand, the fermion pairing gap (blue `+') barely changes during the process, even deep in the insulating side. The pairing-induced insulating phase with $\Delta_b < \Delta_f$, which we are mostly interested in, is clearly present. The small pairing case ($\Delta=0.5$, Fig.~\ref{Fig: 1D fermion boson gap}(b)) shows the same physics. Note that the boson gap is still well-below the fermion gap even when the bare repulsion $U$ is much larger than the fermion gap, because the weakly bound renormalized boson feels a much smaller effective repulsion.  Theoretically, we know the renormalized boson goes through a KT transition at zero temperature in 1+1 dimension. We found the critical U to be around $-0.7$ for $\Delta = 1.3$, and $-0.2$ for $\Delta = 0.5$.

\begin{figure}[htb]
\begin{center}
\includegraphics[width=3in]{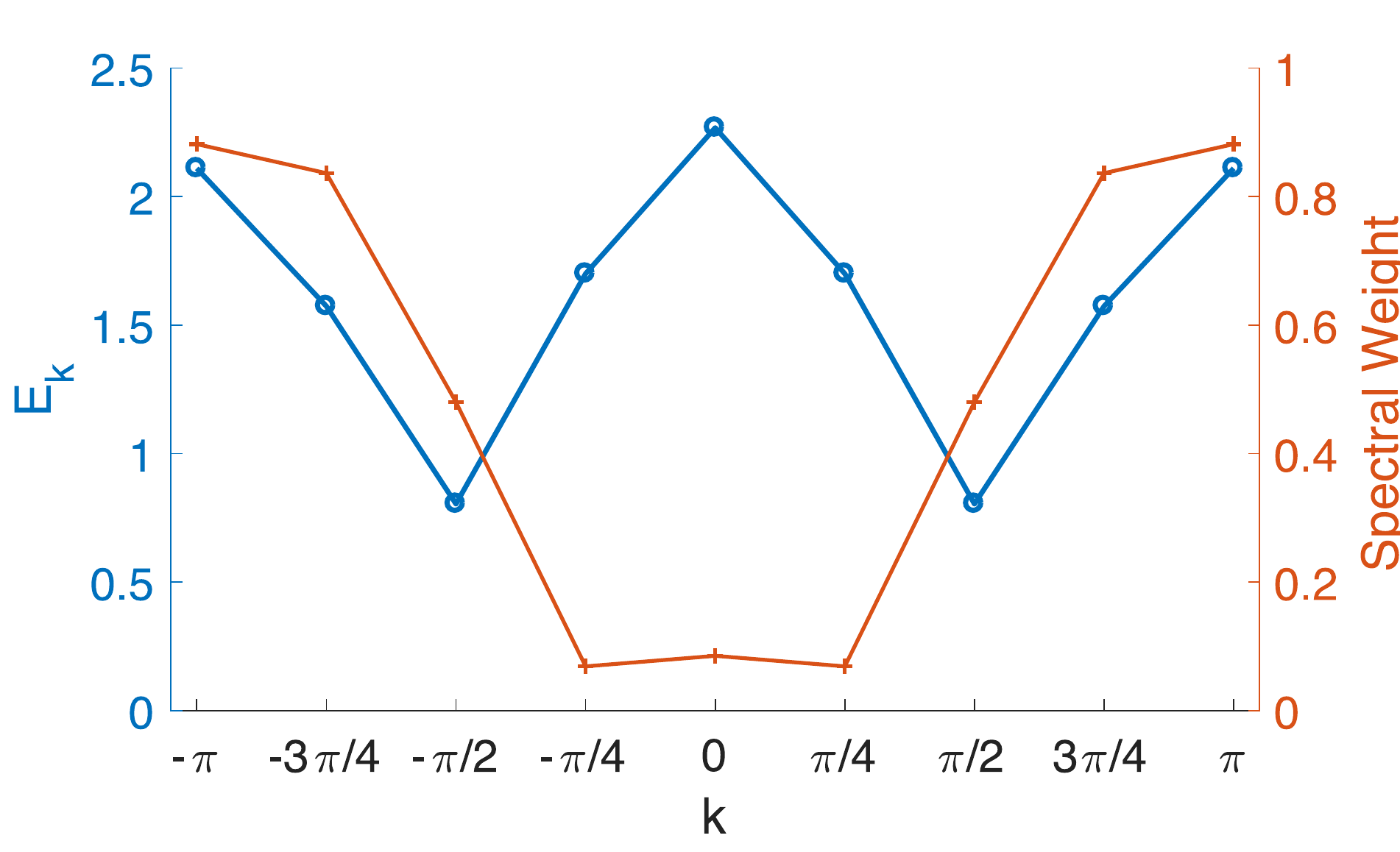}
\caption{Spectrum (blue circles) and spectral weight (red `+') of the lowest charge +1 fermionic excitations, for $-\pi<k<\pi$, $t_b = t_c =1, \D = 1.3, U=-1$. As sketched in Fig.~\ref{Fig: fluctuating SC band}(b), the threshold of fermionic excitations roughly follows the Bogoliubov band. Fermion excitations outside the Fermi sea are quasiparticles. Inside Fermi sea, the thresholds represent 2-particle continuum with zero quasiparticle weight.}
\label{Fig: 1D fermion spectrum weight}
\end{center}
\end{figure}

Finally, we compute the energy threshold for charge-1 excitations at each momentum for $L=8$ (Fig.~\ref{Fig: 1D fermion spectrum weight}) by the Lanczos algorithm. The blue line shows its dispersion, which roughly follows the BCS curve. The red line shows the spectral weight of the excitation: $Z\equiv |\<n|c^{\dagger}_{k}|0\>|^2$. This confirms our physical picture as we illustrated in Fig.~\ref{Fig: fluctuating SC band}. We find that the state for the addition of a single fermion  has considerable overlap with the original fermion for $k>\pi/2$, where the free-fermion band is unoccupied; and vanishing overlap with the original fermion for $k<\pi/2$, where the excitation is essentially hole plus pair.

\subsection{Gapless sector: Fermi pocket}\label{subsection: gapless sector}

\begin{figure*}[htb]
\begin{center}
\includegraphics[width=0.6\linewidth]{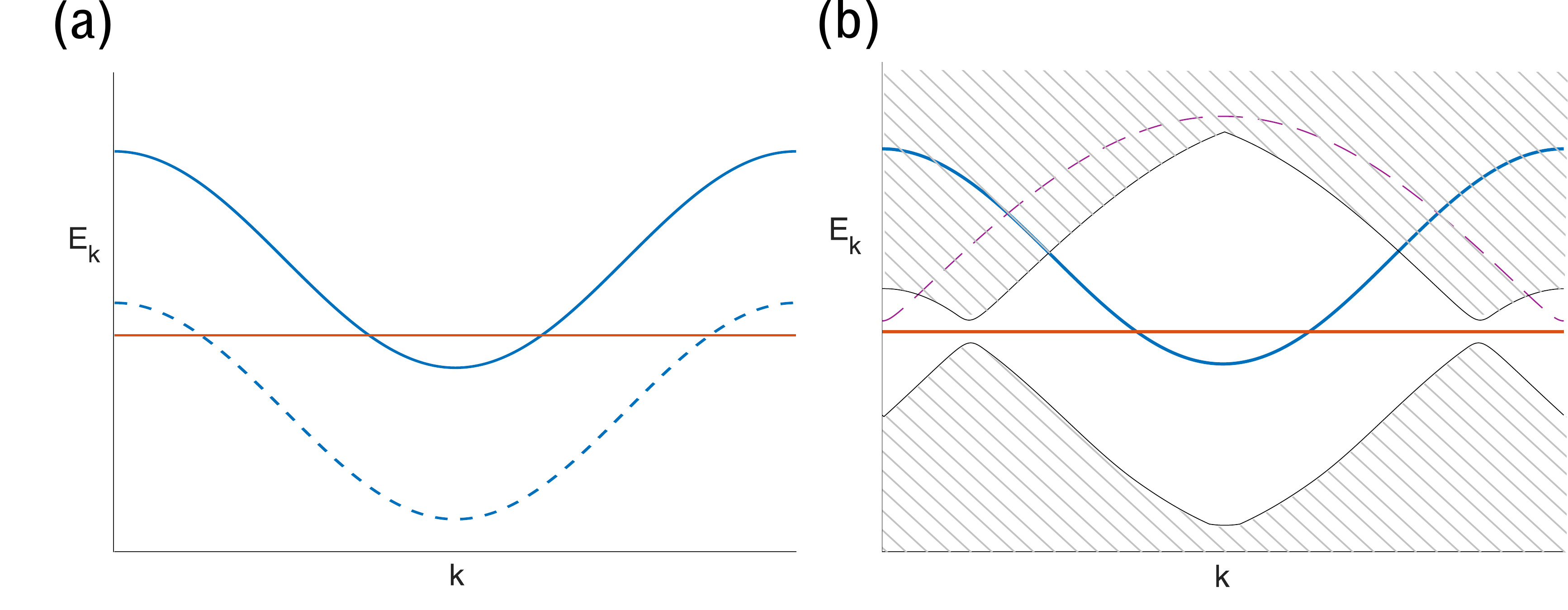}
\caption{Gapless band in (a) PDW-ordered, (b) PDW-disordered state. Solid blue lines in (a) and (b) represent the bare electron dispersion. Solid orange lines in (a) and (b) represent Fermi energy. The dashed blue line in (a) represents the PDW-reflected band. The dashed purple line in (b) represents boson dispersion. The upper/lower shaded area in (b) represents 2-particle continuum of charge $\pm1$, which is calculated from the assumed fermion dispersion (solid blue curve) and boson dispersion (dashed purple curve).}
\label{Fig: gapless band continuum}
\end{center}
\end{figure*}

In the previous two subsections, we use simple models to illustrate the physics relevant to the gapped sector of the fluctuating PDW. We introduce the low-energy effective theory, the boson theory, of the quantum-disordered superconductor, and analyze the influence of the small-gap boson on gapped electrons. 

In this subsection, we use the following model to illustrate the physics of the gapless sector in the fluctuating PDW state.

\bea
H &=& \sum_{k}\e_k c_k^{\dagger}c_{k} + \sum_{k} (E^b_k)^2|\phi_{k}|^2  \nonumber\\
&+& \lambda\sum_{k,q}\phi_{\pi+q}c_{k\uparrow}c_{\pi -k-q\downarrow} + h.c.\\
&=& \sum_{k}\e_k c_k^{\dagger}c_{k} + \sum_{k} E^b_k (b_k^{\dagger}b_k + a_k^{\dagger}a_k)  \nonumber\\
&+& \lambda\sum_{k,q}\frac{1}{\sqrt{E^b_p}}(a_{\pi+q} + b_{\pi-q}^{\dagger})c_{k\uparrow}c_{\pi -k-q\downarrow} + h.c.,
\eea
where $\phi(p) = \frac{1}{\sqrt{E^b_p}}(a_p + b_{-p}^{\dagger})$ is the relativistic boson field describing fluctuating PDW, as introduced in Sec.~\ref{subsection: gapped sector}.
We assume the bare electron has a small pocket at the center of the B.Z., with a dispersion of the solid blue curve in Fig.~\ref{Fig: gapless band continuum}. The bosonic pair ($b_k$) and vacancy of pair ($a_k$) are related by approximate particle-hole symmetry near its superconductor-insulator transition. We assume their band minimum is at momentum $\pi$. We also assume their dispersion is given by the dashed purple curve in Fig.~\ref{Fig: gapless band continuum}(b). In the third term, we are interested in small $q$, and those $k$ around 0 and $\pi$.

If the boson condense at $\pi$-momentum, $\<\phi_\pi\>\equiv\phi_{s}\neq 0$, we can rewrite the fermion in Numbu basis, $\Psi_k\equiv (c_k,c_{\pi - k}^{\dagger})^{\text{T}}$. At the mean-field level

\be
H_f = \sum_k \Psi_k^{\dagger}
\left( \begin{array}{cc}
\e_k & \lambda\phi_s^*\\
\lambda\phi_s & -\e_{\pi - k}
\end{array} \right) \Psi_k
\ee
Since $\e_k$ and $-\e_{\pi - k}$ always have a large difference (Fig.~\ref{Fig: gapless band continuum}(a), solid blue line and dashed blue line), the coupling barely does anything. The band structure is the original electron band plus the reflected band. Due to the small mixing between the two bands,  the new gapless pocket at $\pi$ gains a small electron weight.

If the boson disorders, to the first order, the coupling can be ignored and the fermion maintains its bare single-band dispersion, with only one gapless pocket (Fig.~\ref{Fig: gapless band continuum}(b)). However, the reflected band maintains its presence at finite energy. We can create a hole of the solid blue band and a pair in the dashed purple band to make a 2-particle continuum for electronic excitation. The energy of the two-particle excitation at momentum $k$ can be $|\e_q| + E^b_{k-q}$ for every momentum $q$ such that $\e_q <0$ (so that we can excite a hole at momentum $q$). We calculated possible values of the two-particle excitation energy from the assumed boson and fermion distribution, and illustrate them as the shaded region in the upper half plane. Similarly, there is a two-particle continuum of an electron and a vacancy of pair. The two-particle continuum is strictly gapped since the boson is gapped. When $\D_b$ is small, part of the threshold of the continuum roughly resembles the reflected band shown in Fig.~\ref{Fig: gapless band continuum}(a). The rest of the threshold follows the boson dispersion.

\subsection{Fluctuating PDW in cuprates}\label{subsection: fluctuating PDW in cuprates}

Now we go back to the fluctuating PDW in cuprates. Under the assumption that the pseudo gap is a fluctuating PDW gap, we estimate relevant energy scales as follows. The anti-nodal fermion gap in Bi2212 near 12\% doping, measured by ARPES and STM, is around $60\ $meV. We identify it with $\Delta_f$ in previous theoretical analysis. As we move to the nodal direction, the fermion gap decreases. From the mean field calculation, the lowest gapped band has a gap around $30$ meV. The boson gap has not been measured yet, and we roughly estimate it as follows. Without other obvious velocity scale, we assume the boson velocity to be similar to the \textit{anti-nodal} Fermi velocity. Therefore  $\Delta_b \sim \Delta_f\cdot(\ \text{coherence length}\ /\ \text{correlation length})$, which is between $10$ meV and $30$ meV.

Of the 36 bands (18 pairs of bands) in the mean-field PDW ansatz, 2 are gapless. In the MDW reduced B.Z., the PDW momentum is $(\pi,\pi)$. We apply the theory in Sec.~\ref{subsection: gapless sector} to the gapless bands. After disordering the PDW, the 2 Bogoliubov bands become 1 gapless electron band plus 1 gapped electron-boson continuum. As we discussed in Sec.~\ref{Sec: big picture}, the Fermi pocket automatically adjust its area to satisfy Luttinger's theorem, in order to avoid paying the Mott gap of the bosonic sector. On the other hand, the 34 gapped bands are more complicated than the simple model we have in Sec.~\ref{subsection: gapped sector}. The difference is the existence of many low-lying gapped bands. Thus even though the boson gap is smaller than the anti-nodal gap, it may be larger than the gap of low-lying electrons. However, the picture that  all these fermions are gapped
and that at low enough energy, the bosonic pairs carry all the charges of the  gapped bands is unchanged. At the energy scale of $20$meV, we start to see both fermionic excitations that break pairs and bosonic excitations that move the pair as a whole. Similar to the fluctuating s wave superconductor discussed in Sec.~\ref{subsection: gapped sector}, as we disorder PDW, a Bogoliubov band of ordered PDW evolves into quasi-electron band in part of the B.Z. and  hole-pair continuum elsewhere. Roughly speaking, the Bogoliubov bands coming from the original electron bands become quasi-electron excitation with a 3-particle continuum at slightly higher energy; the Bogoliubov bands coming from PDW-reflected bands become a broad 2-particle continuum with no well-defined quasi-particle (Fig.~\ref{Fig: ARPES th}(a)). This dichotomy is too crude if a large number of bands have similar energy. Generically, the single-particle Green's function mixes multi-boson-fermion contributions from the boson band and all of the fermion bands. Due to the low-energy boson, low-energy two-particle continuum is abundant in the B.Z.

Due to the coexistence of the gapped and gapless sector, and the presence of many low-lying gapped fermion bands, the quasi-particles we discussed previously may be considerably broadened. First, we discuss the fate of the boson. The boson near the PDW momentum cannot decay into the nodal gapless band because of momentum mismatch, otherwise the gapless band would be gapped by PDW in the first place; nor can it decay into the anti-nodal fermions if its energy is smaller than the anti-nodal gap. However, the boson may decay into low-lying gapped fermions: their energy gaps could be comparable (depending on details of the band structure), and the momenta of low-lying fermions cover the majority of the reduced B.Z.. However, the decaying rate should be parametrically small because it relies on the small CDW amplitudes to match the momentum. Thus, even though the boson may not have infinite lifetime, they may still be sharp excitations near the PDW momentum. Second, for the fate of the anti-nodal fermions, since it has a large gap, apart from the boson-fermion continuum we discussed before, the quasi-particle peak itself is also severely broadened by decaying into 3 gapless/small-gap fermions. We shall analyze these spectral features with ARPES and infrared absorption data in the next section.

\section{Broader aspects and experimental implications}\label{Sec: broader aspects experiments}

So far, we have been focusing on the high-field ground state of underdoped cuprates. However, the phenomena we discussed, including the anti-nodal fermion gap, the decrease of fermionic carrier density, and the nodal gapless fermions are also present in the zero-field pseudogap. 
In the limit that the pseudogap transition temperature $T^*\gg T_{c}$ (the superconducting transition temperature), which is achieved in a range of doping, the superconducting phase occupies only a small region of the temperature-field phase diagram, on top of the pseudogap phenomena. In that limit, it is reasonable to expect the pseudogap physics at temperature $T_\text{c}<T\ll T^*$ connects smoothly to the zero-temperature, $H > H_c$ pseudogap ground state we present. Therefore we also compare our theoretical predictions with zero-field finite-temperature data. 

Many finite-frequency spectral properties of the pseudogap is maintained below $\text{T}_\text{c}$. For these properties, we may still use the predictions of our boson-fermion model. However, approaching $\text{T}^*$, the system crosses over to the strange-metal region, where our model does not apply.

On the other hand, it is interesting to discuss fluctuating zero-momentum superconductivity (SC) and fluctuating PDW in a unified picture, and compare their properties. As discussed before, we model the system as nodal electron pocket plus antinodal gapped excitations effectively described by bosonic pairs. The bosonic pair has a local band minimum at finite momentum, which we identified as fluctuating PDW. At low magnetic field and low temperature, cuprates become d-wave superconductors; therefore, the bosonic pair should have another local band minimum at zero-momentum, which closes at $\text{T}_\text{c}$ to give the superconductivity. In the normal state, the 2 band minima of the bosonic Mott insulator give fluctuating PDW and fluctuating SC correspondingly.

The fluctuating SC associated with zero-momentum boson differs from the fluctuating PDW in many aspects. Since it actually orders below $\text{T}_\text{c}$, its fluctuation depends sensitively on temperature. As the first approximation, we may ignore the quantum fluctuation of zero-momentum boson and describe the thermal fluctuation by classical statistical mechanics. On the contrary, since the PDW boson maintains a finite gap everywhere in the phase diagram, thermal fluctuations are largely suppressed. Moreover, the zero-momentum boson decays into the gapless nodal pocket in the normal state, resulting in a considerable dissipation, whereas the PDW boson is immune from that decaying channel and stays relatively sharp because of momentum mismatch. Our discussion on the quantum fluctuation of the PDW is very different from the conventional dissipative Ginzburg-Landau formulation. In that formulation, pairing correlator decays exponentially in real time due to dissipation, $\<\D^*(r,t)\D(r,0)\>\sim e^{-t/\tau}$. However, pairing correlator at the same location oscillates in time in our model, $\<\D^*(r,t)\D(r,0)\>\sim e^{i\D_b t}/t$, with negligible exponential decaying at low temperature, just as every gapped bosonic system. Due to this difference, fluctuating SC, which is close to the conventional thermal fluctuation, produces large Nernst signal and diamagnetism, while the fluctuating PDW boson gives sharper features in spectroscopic measurements. We would like to point out here that the correlator  $\<\D^*(r,t)\D(r,0)\>$ is in principle measurable by tunneling experiments, and a concrete scheme has recently been proposed~\cite{PhysRevB.99.035132}.

Both fluctuating SC and fluctuating PDW modify the spectral function of electrons. On the gapless PDW pocket, the superconducting gap is purely due to d-wave SC; near the antinode, their effects mix together. The combined effect depends on the relative strength of the two, which varies with chemical formula, temperature, and momentum. When $\text{T}^*\gg\text{T}_\text{c}$, we expect the anti-nodal gap to come mainly from fluctuating PDW. Below $\text{T}_\text{c}$, ordered superconductivity gaps out low-lying fermions, hence the reduction of decaying channel for anti-nodal fermions, and the emergence of a sharper anti-nodal peak. As discussed below, this picture is consistent with the data on the single layer Bi2201. On the other hand, for Bi2212 close to optimal doping (still underdoped), a sharp quasiparticle peak emerges from a relatively broad continuum just below $T_c$, and the spectral weight of the peak is apparently proportional to the superfluid density~\cite{feng2000signature,PhysRevLett.87.227001}. This behavior cannot be explained by the fluctuating PDW alone. We also notice that we do not have a clear separation of scale in this situation: $T^*$ is only two times $T_c$. We leave further discussion of Bi2212 to future works.

Underdoped Bi2201, consists of single $\text{CuO}_2$ layers separated far away from each other, has $\text{T}^*$ much bigger than $\text{T}_\text{c}$. It is ideal for analyzing pseudogap effect due to the lack of interlayer splitting and large separation between $\text{T}^*$ and $\text{T}_\text{c}$~\cite{hashimoto2014energy,he2011single}. It has the fermion spectrum closest to what we expect from fluctuating PDW alone. We discuss it in Sec.~\ref{subsection: ARPES}. For other spectroscopic probes, like infrared conductivity and density-density response, we expect to see contributions from fluctuating PDW at $\omega > 2\D_b \sim 40$meV, and contributions from SC at lower frequencies (Sec.~\ref{subsection: infrared}). 

Both fluctuating SC and fluctuating PDW contribute to diamagnetism and Nernst effect. It is well known that as temperature approaches $\text{T}_\text{c}$, the diamagnetism and Nernst signal from fluctuating SC diverges~\cite{PhysRevLett.95.247002,PhysRevLett.88.257003,PhysRevB.73.024510,PhysRevLett.89.287001,larkin2008fluctuation,PhysRevLett.96.147003,PhysRevLett.99.117004,PhysRevB.73.094503}. In contrast, the fluctuating PDW contributions are far less dramatic unless the corresponding boson gap decreases substantially in high fields.

In the following parts of this section, we use our boson-fermion model to work out signatures of the fluctuating PDW. We compare theoretical results with experiments on ARPES, infrared absorption, density-density response, diamagnetism and Nernst effect.

\subsection{ARPES}\label{subsection: ARPES}

As we discussed in Sec.~\ref{subsection: gapped sector} and Sec.~\ref{subsection: fluctuating PDW in cuprates}, the fluctuating PDW state naturally has both charge $\pm2$e bosons and charge $\pm$e electrons/holes at low energy. Their interplay produce unconventional ARPES signal. Since the charge $\pm2$e boson is cheap, when we kick out an electron from the sample, the hole may decay into a charge -2e boson and a charge e electron. In analogy to Fig.~\ref{Fig: fluctuating SC band}(b), the threshold to create a hole excitation at momentum $k$ roughly follows the Bogoliubov bands of PDW, but only in a part of the B.Z. the threshold corresponds to quasi-hole excitations. The other part of the Bogoliubov bands, which comes mainly from PDW reflection, is replaced by a blurred 2-particle continuum of an electron and a small-gap charge -2e boson. Furthermore, wherever we have a sharp quasi-particle in the spectrum, we can add a charge +2e boson and a charge -2e boson to make a 3-particle continuum with the same charge, at the same momentum, and with energy only 2$\Delta_b$ higher. The spectral features of these multi-particle continuum with total charge $-e$, which can be probed by ARPES, are easily calculated by considering the decay rate (the imaginary part of the self-energy) using Fermi's Golden rule or simple dimensional analysis. Consider the simplest coupling $\delta H_1 =\lambda_1 \phi c c + h.c.$ and $\delta H_2 =\lambda_2 \phi^*\phi c^{\dagger}c$, where $\phi(p) = \frac{1}{\sqrt{E^b_p}}(a_p + b_{-p}^{\dagger})$ is the relativistic boson field (see Sec.~\ref{subsection: gapped sector}), with momenta close to the PDW momentum $P$, and $E^b_p = \sqrt{|v_b(p-P)|^2 + \Delta_b^2}$.

\bea
\label{Eq: 2 particle continuum}\text{Im}\Sigma_{2p}(q,\omega) &\propto& \int \dbar^2 p \ \frac{1}{E^b_p}\delta(\omega - E^b_p -E^e_{q-p})\\
&\propto& \theta (\omega - \Delta^{(2)}_q)
\eea
\bea
\text{Im}\Sigma_{3p}(q,\omega) &\propto& \int \ \frac{\dbar^2 p_1\ \dbar^2 p_2}{E^b_{p_1}E^b_{p_2}}\delta(\omega - E^b_{p_1} - E^b_{p_2} -E^h_{q-p_1-p_2})\nonumber\\
\label{Eq: 3 particle continuum}&\propto& (\omega - \Delta^{(3)}_q)\ \theta(\omega - \Delta^{(3)}_q),
\eea
when $\omega - \Delta^{(3)}_q \gg \Delta_b$.

We use the shorthand $\dbar^2 p\equiv \frac{dp_xdp_y}{(2\pi)^2}$. $E^e_{k}/E^h_k$ represents the dispersion of the quasi-electron/ quasi-hole. $\D^{(2)}_q$ ($\D^{(3)}_q$) is the energy threshold to create 2(3) particles at momentum q: $\D^{(2)}_q\equiv \text{min}_{p_1}[E^b_{p_1} + E^e_{q-p_1}]$, $\D^{(3)}_q\equiv \text{min}_{p_1,p_2}[E^b_{p_1} + E^b_{p_2} + E^h_{q-p_1-p_2}]$. When the boson gap is small, and the boson velocity is comparable to the Fermi velocity near the antinode, $\D^{(2)}_q$ and $\D^{(3)}_q$ roughly follows the Bogoliubov bands of PDW.

The main message is that whenever we have a PDW reflected band, we should see a step function in spectral function (Eq.~\ref{Eq: 2 particle continuum}); and whenever we have a (PDW-modified) quasi-hole, we should see a spectral function 

\bea
A(\omega) = \text{Im} \frac{1}{\omega - E^h_q - i(\omega - \Delta^{(3)}_q)\ \theta(\omega - \Delta^{(3)}_q) - i\Gamma},\ \ \ 
\eea
which has a quasi-hole peak together with a 3-particle continuum (Eq.~\ref{Eq: 3 particle continuum}). The spectral signature is a relatively sharp onset of peak at $E^h_q$, but a long $1/(\omega - \Delta_q^{(3)})$ tail above the 3-particle threshold. 

When $\D_b$ is small, $\D^{(3)}_q\simeq E^h_q$, the quasi-hole peak merges with the 3-particle continuum, and

\bea 
A(\omega) \sim \frac{\theta(\omega - E^h_q)}{\omega - E^h_q}
\eea

\begin{figure*}[htb]
\begin{center}
\includegraphics[width=0.7\linewidth]{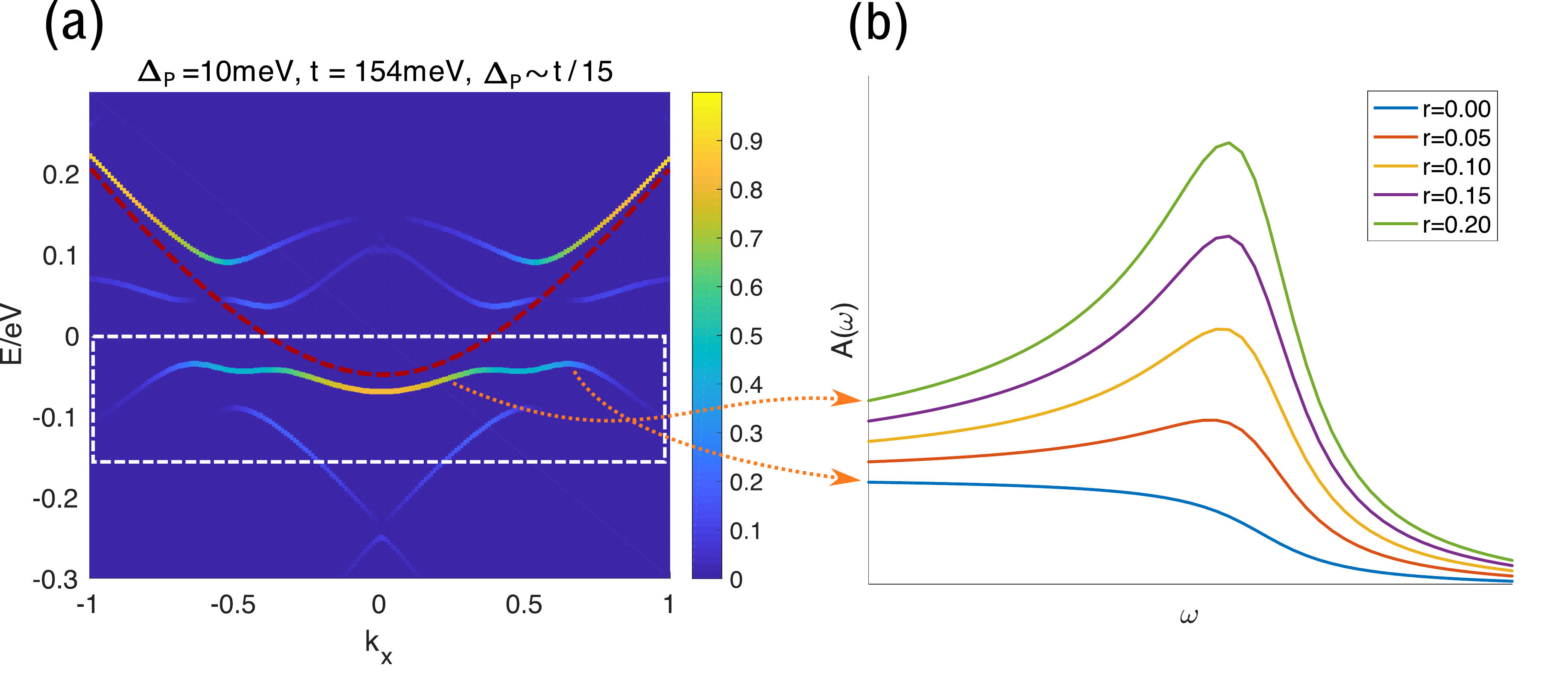}
\caption{(a) Mean-field PDW spectrum along the line $ky=\pi$. PDW momentum $2\pi/6$, PDW pairing $\Delta_P = 10\text{meV}$. (see Eq.~\ref{Eq: PDW mean field} and Eq.~\ref{Eq: PDW form factor} for definition). We use tight-binding band with $t=154\text{meV}, t_p = -24\text{meV}, t_{pp} = 25\text{meV}, t_{ppp} = -5\text{meV}$, chemical potential $\mu = -126\text{meV}$. Color plot represents the spectral weight in mean-field calculation. The dashed red line illustrates the original electron band. The dashed white box shows the range of energy probed by ARPES in Ref.~\cite{he2011single}.  (b) Illustration of the evolution from a 3-particle continuum  to a broad 2-particle continuum of the fluctuating PDW.  We use the ratio $r$ to interpolate  between the two as defined in  Eq.~\ref{Eq: step plus peak}.}
\label{Fig: ARPES th}
\end{center}
\end{figure*}

\begin{figure}[htb]
\begin{center}
\includegraphics[width=0.95\linewidth]{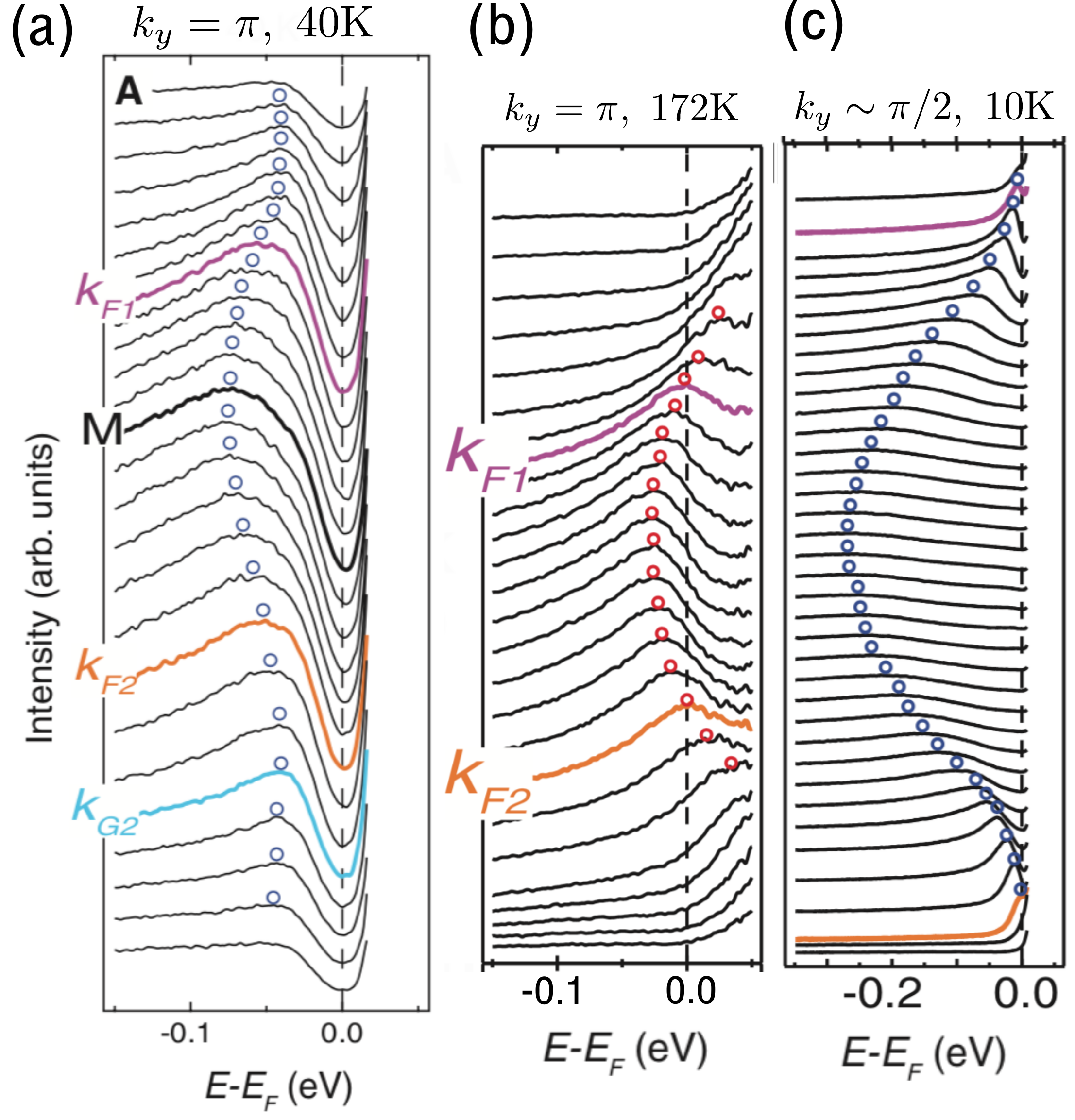}
\caption{(a) Fig. 4A of Ref.~\cite{he2011single}. spectral function along the cut $k_y = \pi$, below $T^*$ and above $T_\text{c}$ (40K). The M point refers to $k_x=0$, $k_{F1}$ and $k_{F2}$ are roughly at $k_x=-0.2 \pi$ and $0.2 \pi$ respectively. (b) Fig. 2A of Ref.~\cite{he2011single}. The same as (b), except at temperature above $T^*$ (172K). (c) Fig. 2N of Ref.~\cite{he2011single}. spectral function along a cut $k_y \sim\pi/2$, at 10K.}
\label{Fig: ARPES exp}
\end{center}
\end{figure}

It's important to know whether ARPES can resolve the boson gap. In Sec.~\ref{subsection: fluctuating PDW in cuprates}, we estimate the boson gap to be $10$meV to $30$meV from the correlation length of PDW. The state of art synchrotron ARPES has an energy resolution of a meV, which can in principle resolve the boson gap. However, the anti-nodal quasi-electron peak is at high energy, suffers from substantial broadening through the process of decaying into gapless/small-gap fermions. When the broadening of quasi-electron peak is comparable $\D_b$, the single-particle peak merges with the 3-particle continuum. We just see a broadened $\theta(\omega - E^h_q)/(\omega-E^h_q)$ peak, as if the boson is gapless.

Fig.~\ref{Fig: ARPES th}(a) shows the mean-field spectrum of bidirectional-PDW with relatively small PDW gap, along the cut $k_y = \pi$. To compare with ARPES results (Fig.~\ref{Fig: ARPES exp}(a), reproduced from Ref.~\cite{he2011single}), we focus on the energy-momentum range in the white box, where the mean-field spectral weight  concentrates on a single Bogoliubov band. Comparing with Fig.~\ref{Fig: PDW 3-band illustration}, we find that a simple 2-band calculation with only y-directional PDW captures main features in this energy-momentum range. This is in contrast with the discussion in ~\cite{PhysRevX.4.031017} which focused on the x-directional PDW. Here we find that the x-directional PDW helps increase the band gap, and produce a flat shoulder near the band minimum.

The sharp spectral function in the mean-field calculation is greatly transformed by the PDW fluctuation. For $k_x < k_F$, the Bogoliubov band follows the original electron band (dashed red line). We expect a broadened $\theta(\omega - E^h_q)/(\omega-E^h_q)$ peak just above the quasi-particle energy. (green line in Fig.~\ref{Fig: ARPES th}(b)). At large $k_x$, the Bogoliubov band is far from the original band of the metal; it largely comes from PDW-reflected bands, which we expect to be a 2-particle continuum when PDW is fluctuating, consequently a (broadened) step function in ARPES. (blue line in Fig.~\ref{Fig: ARPES th}(b)).  Going from small $k_x$ to large $k_x$, we expect the hole excitation created by ARPES to gradually mix with boson-electron bound state, until some $k>k_F$, where the boson and electron no longer bound together. The spectral feature is that a quasiparticle resonance disappears (from the green line to blue line in Fig.~\ref{Fig: ARPES th}(b)) right at the onset of the step-function.

Phenomenologically, we can write the electron annihilation operator as
\bea 
c_k = r_1\tilde{c}_k + r_2\sum_{q}\phi^*_{q}\tilde{c}^{\dag}_{k-q}
\eea 
The first term produces a broad quasi-hole resonance $\theta(\omega - E^h_q)/(\omega-E^h_q)$, and the second term produces a step-function background $\theta(\omega - E^h_q)$. Just to illustrate the qualitative trends, we plot (lorentzian broadened)
\bea
\label{Eq: step plus peak}A(\omega)\propto \theta(\omega - \e_q) + r\theta(\omega - \e_q)/(\omega-\e_q)
\eea
where $r\equiv r_1/r_2$, with gradually increasing $r$ in Fig.~\ref{Fig: ARPES th}(b). In general, $r_1$ and $r_2$ depends on energy and momentum. We know qualitatively how they changes, but near the antinode, we have no reliable way to calculate their energy-momentum dependence. However, when $k\gg k_F$, in the limit $E^0_k\gg \omega, E^h_k$, where $E^0_k$ is the dispersion of the original band without PDW (dashed red line in Fig.~\ref{Fig: ARPES th}(a)), we can treat PDW perturbatively, and the spectral function from the 2-particle continuum is given by

\bea
A(\omega) &\sim& \text{Im}\frac{1}{\omega - E^0_k - i|\Delta|\theta(\omega - E^0_k)}\nonumber\\
&\sim& \frac{|\Delta|}{(E^0_k)^2}\theta(\omega - \e_k)
\eea
Thus the height of the step function quickly decays as we move farther away from $k_F$.

Experimental results along the same cut in Bi2201, just above $T_c$, is shown in Fig.~\ref{Fig: ARPES exp}(a)~\cite{he2011single}. Following the peaks of the spectral functions (blue dots), we see the gap minimum is not at the original Fermi surface ($K_{F1}$ and $K_{F2}$), but shifted outward in momentum ($K_{G2}$), consistent with PDW~\cite{PhysRevX.4.031017}. Moreover, the entire frequency dependence of electron spectral function matches with our expectation of the fluctuating PDW (Fig.~\ref{Fig: ARPES th}(b)). As shown in Fig.~\ref{Fig: ARPES exp}(a), when scanning from large $k_x$ to small $k_x$, we first encounter a step function that onsets at about 20meV and when k is less than the Fermi momentum,  a broad resonance emerging just above the step function. This is as expected from the transition from a bound state of boson and electron into a quasi-hole. Identifying the ARPES results with spectral functions of fluctuating PDW, we get an upper bound of the boson gap, $\D_b \lesssim 20$ meV, consistent with our previous estimation.

There are concerns on whether the step-function background in Bi2212 is intrinsic or an artifact of ARPES due to disorder induced scattering that mixes different momenta~\cite{PhysRevB.69.212509}. However, at least in Bi2201, the step-functions we analyzed appear only in the anti-nodal region (for comparison with the nodal region, see Fig.~\ref{Fig: ARPES exp}(c)), and disappear above $T^*$ (Fig.~\ref{Fig: ARPES exp}(b)), providing strong evidence that they are intrinsic and related to the pseudogap. We also notice that these step functions start at around $20$ meV below Fermi energy, different from the step functions that start right at Fermi energy in Bi2212.

Bi2201 is ideal for analyzing the pseudogap for the large separation between $T_c$ and $T^*$ even close to optimal doping, and for the lack of bilayer splitting~\cite{hashimoto2014energy,he2011single}. We found the anti-nodal spectrum of Bi2201 fitted best with a relatively small PDW pairing, $\D\sim t/15$. We also notice that if pairing were to be increased to $\D\sim t/4$, the band structure is no longer captured by a simple 2-band hybridization: there are many bands sharing small spectral weights. Considering PDW fluctuation, the spectral function may just be a featureless continuum above PDW gap. This large-pairing scenario may be the case for other cuprates with larger $T_c$ and $T^*$.

\subsection{Infrared conductivity and density-density response}
\label{subsection: infrared}

\begin{figure*}[htb]
\begin{center}
\includegraphics[width=0.7\linewidth]{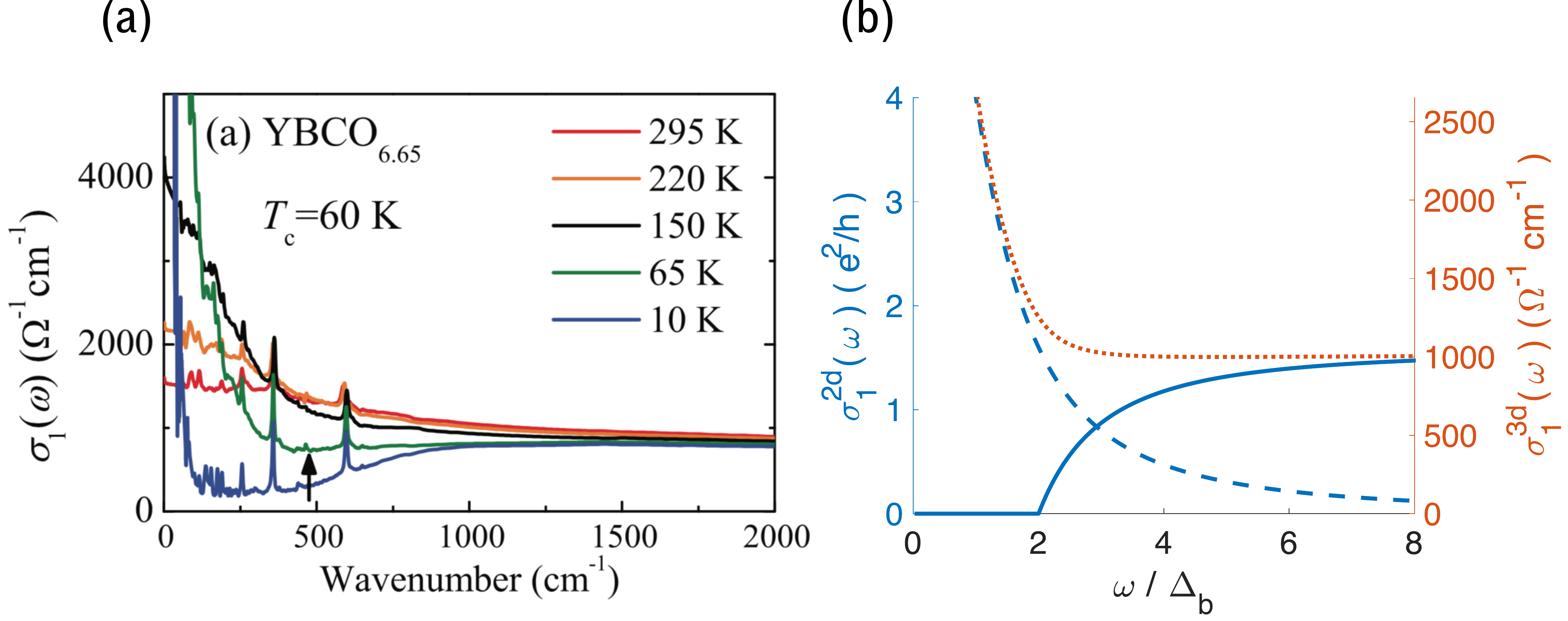}
\caption{(a) Real part of infrared conductivity measured from reflectance (Fig. 3(a) of Ref.~\cite{PhysRevB.90.014503}). (b) Solid blue curve: AC conductivity of a free charge 2e boson with gap $\D_b$. We calculated the 2D conductivity of each layer, and converted it to a 3D conductivity using the lattice parameter of YBCO. The conductivity of the free relativistic boson saturates at $\frac{\pi}{2}e^2/h$ when $\omega\gg\D_b$, which corresponds to $1.0\times10^3\ \Omega^{-1}\text{cm}^{-1}$. Dashed blue curve: a Drude peak. Dashed orange curve: the sum of the boson conductivity (blurred by a Lorentzian) and the Drude peak.}
\label{Fig: infrared}
\end{center}
\end{figure*}

Cuprates have a flat ab-plane infrared conductivity plateau, which differs from a Drude peak that decays as $1/\omega^2$ at high frequencies~\cite{puchkov1996pseudogap,RevModPhys.77.721}. As temperature lowers, the low-frequency peak become narrower, and the conductivity shows an upturn in the infrared region, starting roughly at $40$meV. This extra infrared conductivity have never been throughly understood. Ref.~\cite{puchkov1996pseudogap,PhysRevB.90.014503,PhysRevB.57.R11089,PhysRevLett.81.4716} attempt to explain it by electron scattering with charge-neutral boson. However, we find that it matches well with the conductivity of a charge 2e boson.

Consider a free boson with charge $e^*$, minimally coupled to electromagnetic field.

\bea
\mathcal{L} = \frac{1}{2}|(\partial_{t} + ie^{*}V)\phi|^2 - \frac{1}{2}\sum_{i=1,2}v_b^2|(\partial_{i} + ie^{*}A_i)\phi|^2\nonumber\\ - \frac{1}{2}|\D_b|^2|\phi|^2,\ \ 
\eea

where the momentum of the boson is measured from the PDW momentum. By canonical quantization, $E^b_p = \sqrt{\D_b^2 + v_b^2 p^2}$, $\phi_p = \frac{1}{\sqrt{E^b_p}}(a_{p} +b_{-p}^{\dagger})$, and

\bea
j_{i} = \frac{\delta\mathcal{L}}{\delta A^{i}} = \sum_{p} \frac{e^{*}v_b^2}{E_p} p_{i}(a^{\dagger}_{-p} + b_p)(a_{-p} + b_p^{\dagger})
\eea

By Kubo formula

\bea
\text{Re} \sigma_{xx}(\omega) &=& \frac{\pi}{\omega}\sum_{n}|\<n|j_x|0\>|^2\delta(\omega - (E_n - E_0))\\
&=& \frac{(e^*)^2v_b^4\pi}{\hbar\omega}\int \dbar^2 p\ \frac{p_{x}^2}{(E^b_p)^2}\ \delta(\omega - 2E^b_p)\\
&=& \frac{(e^*)^2}{16\hbar}(1-4\Delta_b^2/\omega^2)\theta(\omega - 2\Delta_b)
\eea

We plot the result for $e^* = 2e$ as the solid blue curve in Fig.~\ref{Fig: infrared}(b), and convert the 2D conductivity to the 3D in-plane conductivity using the lattice constant of YBCO. The optical conductivity of the boson depends on its dispersion and interaction, hence non-universal. However, the linear onset of conductivity at $\omega\simeq 2\D_b$, namely $\sigma_{xx}\propto (\omega - 2\D_b)\theta (\omega - 2\D_b)$, is universal for a gapped boson. The onset is linear because of the combination of a constant density of states in 2D and an absorption matrix element $\sim\text{velocity}^2\sim\delta\omega$, for $\omega\simeq 2\D_b$. For a free relativistic boson with charge 2e, the conductivity at high frequency saturates at $\sigma_{xx} = \frac{\pi}{2}e^2/h$, independent of its gap or velocity. The linear onset of the conductivity together with the saturation value of an order 1 number times $e^2/h$ are signatures of a gapped relativistic particle. Interactions and changes in dispersion modify the order 1 number, but does not change the qualitative features of the conductivity. (For detailed explanation and calculation, see  Ref.~\cite{sachdev_2011}).

Surprisingly, the infrared conductivity plateau around 12\% doping is almost exactly $\frac{\pi}{2}e^2/h$ per $\text{CuO}_2$ layer, the same as the free boson, both in YBCO and in Bi2212. (It changes a little with doping. See Fig.~\ref{Fig: infrared} for comparison with YBCO. See Fig.6 of Ref.~\cite{puchkov1996pseudogap} for Bi2212.) Moreover, the frequency dependence of bosonic conductivity matches well with the conductivity upturn at low temperature. If we add a Drude peak to the bosonic conductivity, we reproduce the flat infrared conductivity observed at higher temperature. 

The extra infrared conductivity provide evidence for the charge 2e boson. However, the numerical agreement may not be taken too seriously, for the interaction between bosons and fermions may modify the result. Experimentally, the infrared plateau extends to frequency as high as $400$meV~\cite{PhysRevB.43.7942}, where our boson fermion model does not apply. We cannot explain the high-energy behavior of the plateau, but we suspect that the boson contribution connects to the incoherent part of the spectral weight (also seen in ARPES) to give the long plateau.

Note that even though the conductivity upturn is prominent only below $T_c$, it has little to do with the absorption across the SC gap. As discussed in Ref.~\cite{PhysRevB.90.014503},  features of SC is around 100$\text{cm}^{-1}\sim 12$meV, five times smaller than the frequency scale of the upturn. Although not fully understood, ordered SC seems to make the low-energy peak narrower without changing the conductivity upturn starting from $40$meV. If we associate the infrared conductivity upturn to the PDW boson, the boson gap should be $20$meV, consistent with our previous estimation.

Unlike s-wave SC, fermions gapped by PDW  absorb light across the pairing gap even in the clean limit. ~\cite{PhysRevB.95.014506}  However, this is much smaller than the bosonic contribution,  according to the estimation in Ref.~\cite{PhysRevB.95.014506}, which found  $\sigma^{2D}_f \sim \frac{e^2}{h} (a/\lambda)^2 E_f/\D_f\sim \frac{1}{10}e^2/h$, where $a$ is size of the original unit cell, $\lambda\sim 8a$ is the wavelength of PDW. The absorption due to  the gapped fermion bands give various tiny peaks from $50$meV to $200$meV, which may be too small to identify. The delta function peaks observed experimentally are mostly due to optical phonons.

The same phenomena is also observed in density-density response. By current conservation, we expect
\bea
\text{Im}\,\Pi(q\sim 0, \omega) &=& \text{Im}\,\<\rho\rho\> = \text{Im}\,\<jj\>\cdot q^2/\omega^2\nonumber\\
&=& \text{Re}\,\sigma(\omega)\cdot q^2/\omega\\
\text{Im}\,\Pi(q\sim 0, \omega) &\sim& \frac{\pi}{2}\frac{e^2}{h}\frac{q^2}{\omega},\text{ in mid-infrared}
\eea
Abbamonte's group measured the density-density response in cuprates~\cite{mitrano2018anomalous,husain2019crossover}. Below 100meV, they claim the signal is dominated by phonon. Between 100meV and 1eV, at optimal doping, they report an unusual $\text{Im}\,\Pi$ independent of $\omega$. In overdoped samples, $\text{Im}\,\Pi$ decreases as $\omega$ decreases to 100meV. However, in underdoped samples, $\text{Im}\,\Pi$ increases as $\omega$ decreases to 100meV~\cite{husain2019crossover}. While this upturn is unusual in metallic states, here it is simply required by current conservation (see Eq. 27) to be consistent with the infrared conductivity.

Finally, we discuss c-axis conductivity. For bilayer cuprates like YBCO and Bi2212, $\text{CuO}_2$ layers are organized as closed bilayers with several atomic layers between neighbouring bilayers. We show in Appendix ~\ref{Appendix: c axis}, that given the experimental fact $\sigma_{zz}\gg \omega\epsilon_0$ in the mid-infrared, the measured conductivity is always determined by inter-bilayer hopping instead of intra-bilayer hopping, as long as we are away from sharp resonances. Physically, the intra-bilayer hopping is so effective that all voltage drop are on the barrier between neighbouring bilayers. Across this barrier of 3 or 4 atomic layers, pair hoping is much smaller than single-fermion hopping. Therefore, we expect tunneling of the small-gap fermion to  dominate the measured c-axis conductivity. For more details, and for the calculation of the bosonic contribution, see Appendx~\ref{Appendix: c axis}.

\subsection{Remnants of superconductivity}

Long-range ordered PDW breaks charge conservation and is a superconducting order. Being close to the long-range PDW, the fluctuating PDW state has properties reminiscent of a superconductor. In this subsection, we briefly discuss the diamagnetic response, Nernst effect, and DC conductivity of the fluctuating PDW state. In short, fluctuating PDW gives a diamagnetic susceptibility inversely proportional to the boson gap without increasing the DC conductivity. This is because the bosons transit from a superconductor into an insulator instead of a metal. Nernst effect comes from thermally excited PDW bosons, which are suppressed when $T<\D_b$. Experimentally observed diamagnetism and Nernst signal near $T_c$ comes mainly from fluctuating zero-momentum SC. Due to the boson gap, the contribution from fluctuating PDW  is smaller and less sensitive to temperature.

We start from diamagnetism. We calculate the current response to the vector potential $j_i(\omega,q) = K_{ij}A_{j}$, at $\omega = 0, q = q_y\hat{y}$. In this setting, magnetic susceptibility of the boson $\chi_b = -K_{xx}/q_y^2$.

The current operator at finite $q$ is
\bea
j_i(q) = \sum_p e^*v^2_b(p_i + q_i/2)\phi^*(p)\phi(-p-q) \nonumber\\
+ \sum_p(e^*)^2v^2_b\phi^*(p)\phi(-p)A_i(q)
\label{Eq: current}
\eea

The response of the first term is given by Kubo formula

\bea
\text{Re} R_{xx} &=& \sum_n|\<n|j_x(q)|0\>|^2\frac{-2}{E_n-E_0}\\
&=& (e^*)^2v_b^4\int_0^{\Lambda}\frac{-2p_x^2\ \dbar^2p}{E^b_p E^b_{p+q}(E^b_p+E^b_{p+q})}
\eea
We expand the expression in $q_y$, the constant term is canceled b{}y the second term of Eq.~\ref{Eq: current}, and the quadratic term gives us magnetic susceptibility

\bea
\chi_b &=& -\text{Re}\frac{R_{xx}(q_y) - R_{xx}(0)}{q_y^2}\\
&=& -(e^*)^2v_b^4 \int\frac{p_x^2}{(E^b_p)^3}\left(-\frac{5}{2}\frac{v_b^4 p_y^2}{(E^b_p)^4} + \frac{3}{4}\frac{v_b^2}{(E^b_p)^2}\right)\dbar^2p\nonumber\\
\label{Eq: diamag}&=& -\frac{e^2v_b^2}{6\pi\D_b}\\
&=& \chi_f \frac{2mv_b^2}{\Delta_b},
\eea
for $e^*=2e$, where $\chi_f$ =  $e^2/12 \pi m$ stands for Landau diamagnetic susceptibility for 2D free fermion with mass $m$. $\chi_b^{3D} = \chi_b/d$, where $d$ is the average distance between $\text{CuO}_2$ layers. This result holds for temperature and Landau-level splitting smaller than the boson gap. We note that compared with $\chi_f$ 
, Eq.~\ref{Eq: diamag} is enhanced by the ratio $\frac{2mv_b^2}{\D_b}$. There has been report of a significant amount of diamagnetism in underdoped YBCO at low temperatures at 40T magnetic field which is much larger than the transport $H_{c2}$. \cite{yu2016magnetic} Our Eq.~\ref{Eq: diamag} involves the boson velocity $v_b$ which is not known, but the predicted diamagnetism should be temperature dependent on the scale of the boson gap.

When the temperature is comparable to or larger than the boson gap, with external magnetic field, bosons exhibit Nernst effect. Under temperature gradient and magnetic field, thermally excited charge 2e and charge -2e bosons drift in different directions, giving a net electric current. 

For temperature smaller than the boson gap and the lowest fermion gap, and away from the superconducting dome, DC conductivity, Hall conductivity, specific heat and quantum oscillation comes solely from the small electron pocket. This decrease of fermionic carrier density at low energy is the main consequence of the fluctuating PDW. However, it is hard to describe how conductivity changes as we enter the pseudogap region from high temperature, since we do not have a theory for the strange metal. 

\subsection{Symmetry breaking in the pseudopgap phase.}
\label{subsection:Symmetry breaking in the pseudopgap phase.}

\begin{figure}[htb]
\begin{center}
\includegraphics[width=0.6\linewidth]{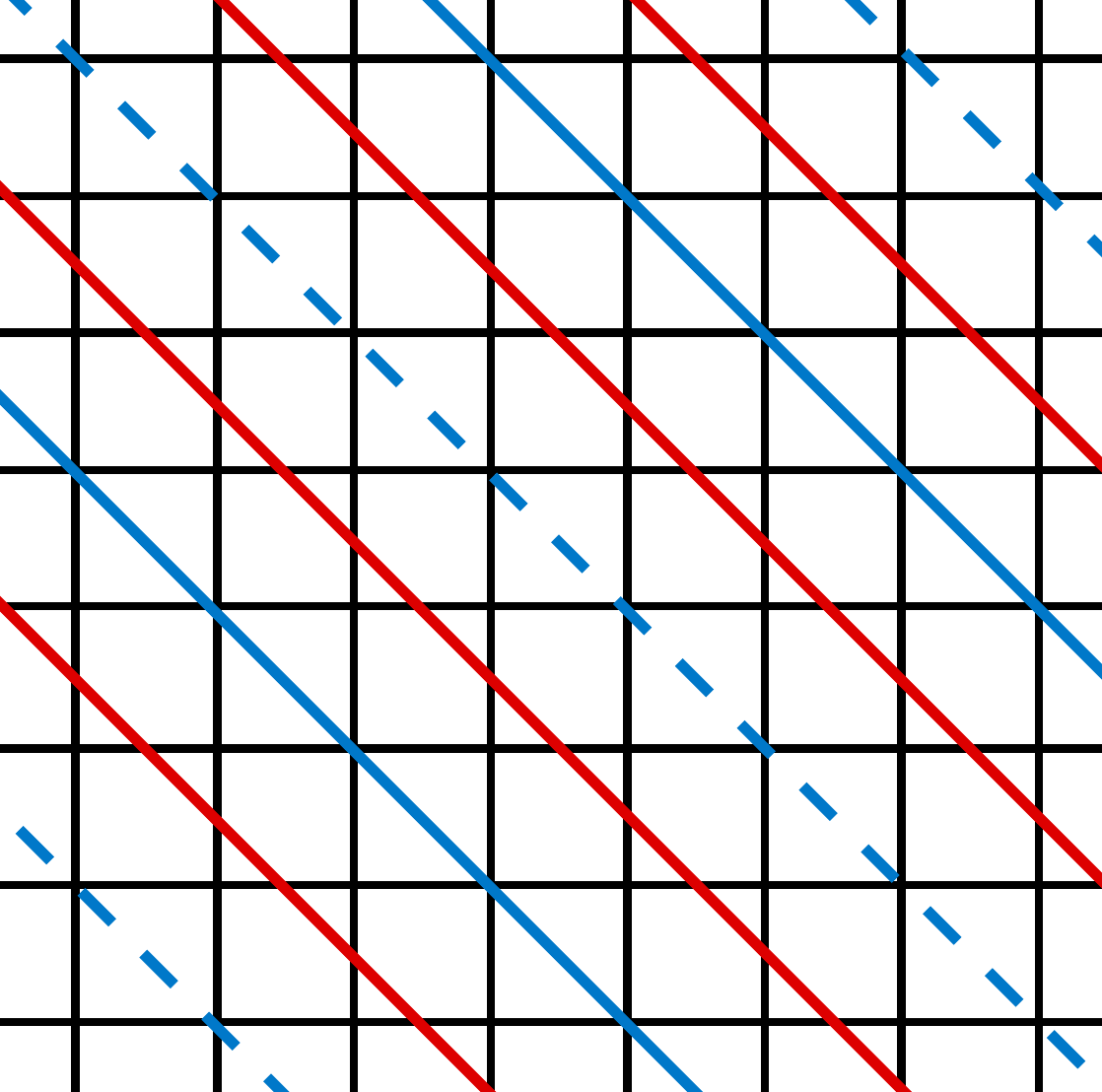}
\caption{Illustration of a uni-directional MDW generated by period-6 PDW. The line of maximum and minimum of the magnetization is shown as solid and dashed blue lines. The zero of magnetization is shown as red lines. Black lines shows the underlying lattice.}
\label{Fig: MDW}
\end{center}
\end{figure}

In this section we consider the consequences of symmetry breaking of the MDW, which is one of the composite orders associated with the PDW. We consider the case of commensurate PDW, and for concreteness we first discuss the case $P=2\pi/6$.  We have many different choices of phases corresponding to different relative positions between the lattice and the CDW/MDW.(see Appendix A for a detailed explanation of these phases.) Lattice translations change PDW phases only by multiples of $2\pi/6$. A generic choice breaks all lattice symmetry, but it may require the CDW/MDW to be pinned at a unnatural position. We focus on the case where the maximum and minimum of a uni-directional MDW at momentum $(P\hat{x},P\hat{y})$ is on site, as shown by the blue lines in Fig.~\ref{Fig: MDW}. 
We shall see that this choice preserves inversion about the origin, but breaks all mirrors perpendicular to the plane. The MDW has magnetization $\vec{M}\propto \cos(Px+Py)\hat{z}$ which breaks mirror symmetry along both (1,1) and (1,-1) since magnetization is odd under mirror. On the other band, we can consider the mirror plane passing through the lines of zero magnetization (shown in red in ~\ref{Fig: MDW}). The mirror symmetry is preserved for the magnetization which is odd in this case, but is broken by the lattice. Thus in this example all mirror planes normal to the c-axis are broken. The same conclusion holds for $P=2\pi/7$. The exception is $P=2\pi/8$ where the line of zero's pass through a lattice site and mirror symmetry is preserved. 

Incommensurate PDWs are slightly more complicated. For the case of YBCO, the PDW wavelengths changes with doping between 6 and 7 lattice spacing. Distorted by lattice, it is natural to relax the cosine waves into domains with period-6 PDW and domains with period-7 PDW. Our discussion of mirror symmetry breaking also applies to this relaxed incommensurate PDW.

So far, we have been focusing on simplified situations where every relative phase between two PDW order parameters are perfectly ordered.  However, as temperature decreases, different relative phases, hence different density waves can order in turn. Although fluctuating PDW gives the tendency of CDW and MDW in both directions, the energy functional may actually prefer a unidirectional MDW/CDW, with a shorter range MDW/CDW in the orthogonal direction, at least in a range of temperature. 
Ref.~\cite{sato2017thermodynamic} reported a nematic phase transition at the onset of the pseudogap. This is most clear in the case of the Hg compound which has a tetragonal structure and the nematicity is along the diagonal. This result may be explained if the MDW preferentially forms short range order at momentum $P\hat{x} + P\hat{y}$  at $T^*$ without the MDW at $P\hat{x}-P\hat{y}$, giving rise to a nematic transition. 

In Ref.~\cite{PhysRevB.97.174511}, we estimated the magnetic moment per plaquette (of the Cooper lattice) is at the order of $5\times 10^{-3}$ Bohr magneton. The moment through a half period of the MDW is larger by the corresponding area and we estimate the magnetic field generated by this moment to be $\sim 0.5$ Gauss. However, the magnetic field changes smoothly in the range of 6 or 7 lattice spacing. In NMR experiments, such a magnetic field profile gives a broadening of the resonance peak, instead of a shift of the peak, therefore hard to detect. But the MDW may be detectable by neutron scattering.

\section{Discussion}

In this paper, we discuss the low-energy effective theory of the pseudogap, relevant for underdoped cuprates when $T^*>>T_c$, and for the high-field ground state. We disorder bidirectional pair density waves, but maintaining the descendant orbital magnetization and charge density waves to get a ground state of small electron pocket and a hidden bosonic Mott insulator. The fluctuating PDW provides a smooth background for diamagnetism and Nernst effect on top of fluctuating zero-momentum superconductivity, without producing excess DC conductivity. We present detailed comparison of the theoretical predictions and the experiments on ARPES and infrared conductivity. We found the peculiar spectroscopic features of the pseudogap is consistent with having a small-gap charge 2e boson at finite momentum, as in our proposal for the fluctuating PDW. From the measured infrared conductivity and the correlation length of PDW in the vortex halo, we estimate the boson gap to be about $20$meV. However, infrared conductivity and ARPES probes only the two particle continuum of two bosons or of a boson and an electron. A direct probe of a single charge 2e boson near $20$meV, momentum $2\pi/8\sim2\pi/6$ would provide direct evidence for our proposal. We also propose an orbital magnetization density wave in $(1,1)$ direction, with momentum $1/\sqrt{2}$ of the momentum of CDW. This MDW breaks time reversal, and it could explain the nematic transition at the onset of the pseudogap~\cite{sato2017thermodynamic}. We have not discussed how the pseudogap descends from the strange metal, but it would be very interesting to explore the relation between our model and possible theories of the strange metal.

\section{Acknowledgment}
We thank Sudi Chen, Alberto de la Torre, Ruihua He, David Hsieh, Lu Li, N. P. Ong, Joe Orenstein and Z. X. Shen for discussion and helpful information. DMRG calculations were performed using the ITensor Library\cite{ITensor}. TS is supported by a US Department of Energy grant DE- SC0008739, and in part by a Simons Investigator award from the Simons Foundation. PAL acknowledges the  support by DOE office of Basic Sciences grant number DE-FG02-03ER46076.

\bibliographystyle{apsrev4-1}
\bibliography{reference_pseudogap}

\appendix

\section{Symmetry of the fluctuating PDW state}\label{Appendix: Symmetry of the fluctuating PDW state}

Before we discuss the symmetry of fluctuating PDW states, it is helpful to have in mind a specific pairing form factor in real space. We choose a local d-wave form factor. Define

\bea
S[(m,n), (m',n')] = c_{m,n,\uparrow}c_{m',n',\downarrow} - c_{m,n,\downarrow}c_{m',n',\uparrow}\nonumber \\
b_{m,n} = S[(m,n),(m+1,n)] + S[(m,n),(m-1,n)]\nonumber\\
- S[(m,n),(m,n+1)] - S[(m,n),(m,n-1)]\ \ \ 
\eea
where $(m,n)$ labels a Cu site in $\text{CuO}_2$ plane. $S[(m,n),(m',n')]$ represents a singlet pairing between two sites; $b_{m,n}$ represents d-wave pairing on nearest-neighbor bounds. (The following analysis is not restricted to this specific form.) A simple Hamiltonian with 4 PDWs can be

\bea
H = \sum_{m,n}\sum_{\ \vec{p} = P\hat{x}, P\hat{y}, -P\hat{x}, -P\hat{y}}\D_{\vec{p}} \,e^{i\vec{p}\cdot (m,n)} b_{m,n} + h.c.\ \ 
\eea
In order to gain pairing energy from all anti-nodal fermions, and for the approximate $C_4$ symmetry of $\text{CuO}_2$ plane, we assume the 4 PDW amplitudes have approximately equal amplitude. At low temperature, we assume only the overall superconducting phase of the 4 PDW order parameters is fluctuating. Relative phases between every pair of PDW order parameters are all ordered.

Time reversal symmetry maps $(\D_{P\hat{x}}, \D_{P\hat{y}}, \D_{-P\hat{x}}, \D_{-P\hat{y}})$ to $(\D^*_{-P\hat{x}}, \D^*_{-P\hat{y}}, \D^*_{P\hat{x}}, \D^*_{P\hat{y}})$. Time reversal invariance requires that these two set of phases differ only by an overall $U(1)_\text{charge}$ transformation.

\bea
\text{Time reversal: }(\D_{P\hat{x}}, \D_{P\hat{y}}, \D_{-P\hat{x}}, \D_{-P\hat{y}}) =\nonumber\\
e^{i\phi}(\D^*_{-P\hat{x}}, \D^*_{-P\hat{y}}, \D^*_{P\hat{x}}, \D^*_{P\hat{y}})
\eea
Similarly, invariance under inversion (about (0,0)), and Mirror along (1,-1) direction (passing through (0,0)) requires
\bea\label{Eq: inversion}
\text{Inversion about (0,0): }(\D_{P\hat{x}}, \D_{P\hat{y}}, \D_{-P\hat{x}}, \D_{-P\hat{y}}) \nonumber\\
=e^{i\phi'}(\D_{-P\hat{x}}, \D_{-P\hat{y}}, \D_{P\hat{x}}, \D_{P\hat{y}})\ \\
\text{Mirror along (1,-1): }(\D_{P\hat{x}}, \D_{P\hat{y}}, \D_{-P\hat{x}}, \D_{-P\hat{y}}) \nonumber\\
=e^{i\phi''}(\D_{-P\hat{y}}, \D_{-P\hat{x}}, \D_{P\hat{y}}, \D_{P\hat{x}})\ \label{Eq: Mirror}
\eea
where we have chose the mirror passing through $(0,0)$.

Last, under translation, $(x,y)\rightarrow (x,y) + (a,b),\ (a,b)\in\mathbb{R}^2$, 
\bea
(\D_{P\hat{x}}, \D_{P\hat{y}}, \D_{-P\hat{x}}, \D_{-P\hat{y}}) \rightarrow\nonumber\\
(e^{iPa}\D_{P\hat{x}}, e^{iPb}\D_{P\hat{y}}, e^{-iPa}\D_{-P\hat{x}}, e^{-iPb}\D_{-P\hat{y}})
\eea

To the second order of PDW amplitudes, CDW and MDW at momentum $P\hat{x}+P\hat{y}$ are generated:

\bea
\rho_{P\hat{x} + P\hat{y}} = c(\D_{P\hat{x}}\D^*_{-P\hat{y}} + \D_{P\hat{y}}\D^*_{-P\hat{x}}),\ c\in\mathbb{R}\\
M_{P\hat{x} + P\hat{y}} = id(\D_{P\hat{x}}\D^*_{-P\hat{y}} - \D_{P\hat{y}}\D^*_{-P\hat{x}}),\ d\in\mathbb{R}
\eea
where $\rho$ is charge density, $M \equiv \hat{z}\cdot\nabla\times\vec{j}$ is the orbital magnetization in $\hat{z}$ direction. Time reversal symmetry and inversion symmetry of the theory requires $c$ and $d$ to be real and exclude other free parameters. To give an example of this symmetry argument, we analyze the coefficients of MDW. By momentum and charge conservation, and that the magnetization is real in real space, the most general form of MDW at the second order is
\bea
M_{P\hat{x} + P\hat{y}} = d_1\D_{P\hat{x}}\D^*_{-P\hat{y}} + d_2\D_{P\hat{y}}\D^*_{-P\hat{x}}\\
M_{-P\hat{x} - P\hat{y}} = d_2^*\D_{-P\hat{x}}\D^*_{P\hat{y}} + d_1^*\D_{-P\hat{y}}\D^*_{P\hat{x}}
\eea
Consider the time reversal partner of the system, with pairing amplitude
$(\tilde\D_{P\hat{x}}, \tilde\D_{P\hat{y}}, \tilde\D_{-P\hat{x}}, \tilde\D_{-P\hat{y}})$ $= e^{i\phi}(\D^*_{-P\hat{x}}, \D^*_{-P\hat{y}}, \D^*_{P\hat{x}}, \D^*_{P\hat{y}})$.

\bea
\tilde M_{P\hat{x} + P\hat{y}} = d_1\tilde \D_{P\hat{x}}\tilde \D^*_{-P\hat{y}} + d_2\tilde \D_{P\hat{y}}\tilde \D^*_{-P\hat{x}}\nonumber\\
=d_1\D^*_{-P\hat{x}}\D_{P\hat{y}} + d_2\D^*_{-P\hat{y}}\D_{P\hat{x}}\\
\tilde M_{-P\hat{x} - P\hat{y}} = d_2^*\tilde \D_{-P\hat{x}}\tilde \D^*_{P\hat{y}} + d_1^*\tilde \D_{-P\hat{y}}\tilde \D^*_{P\hat{x}}\nonumber\\
= d_2^*\D^*_{P\hat{x}}\D_{-P\hat{y}} + d_1^*\D^*_{P\hat{y}}\D_{-P\hat{x}}
\eea
Since $\tilde M(x) = -M(x)$, we know that $d_1 = -d_2$. Similar arguments for inversion requires $d_1 = d_2^*$. Thus $d_1=-d_2 =id, \ d\in\mathbb{R}$. Similarly, time reversal symmetry of the theory requires the density wave generated in the leading order at momentum $2P\hat{x}$ and $2P\hat{y}$ are pure CDW with no magnetization.

In the limit PDW wavelength is much larger than the lattice spacing, we can use two lattice translation and $U(1)_\text{charge}$ to continuously change 3 of the 4 phases of the PDW amplitudes. In this limit, the only nontrivial phase is

\bea
e^{i\theta} \equiv \frac{\D_{P\hat{y}} \D_{-P\hat{y}} }{\D_{P\hat{x}} \D_{-P\hat{x}} }
\eea
This phase determines whether we have CDW or MDW at momentum $P\hat{x}+P\hat{y}$, and it affects the band structure (Fig.~\ref{Fig: Bogoliubov bands}). Time reversal symmetry forbids MDW, and requires $\theta = 0$, hence a CDW at momentum $P\hat{x}+P\hat{y}$. However, such a CDW is not observed experimentally. We postulate the opposite scenario, $\theta = \pi$, with only MDW at momentum $P\hat{x}+P\hat{y}$, which breaks time reversal. In the long-wavelength limit, inversion symmetry and mirror symmetry are always preserved. We can always find an inversion center and a mirror by translation. In the main text we consider further the case of finite wavevector P.

\section{c-axis conductivity and the boson contribution}
\label{Appendix: c axis}

\begin{figure}[htb]
\begin{center}
\includegraphics[width=2.5in]{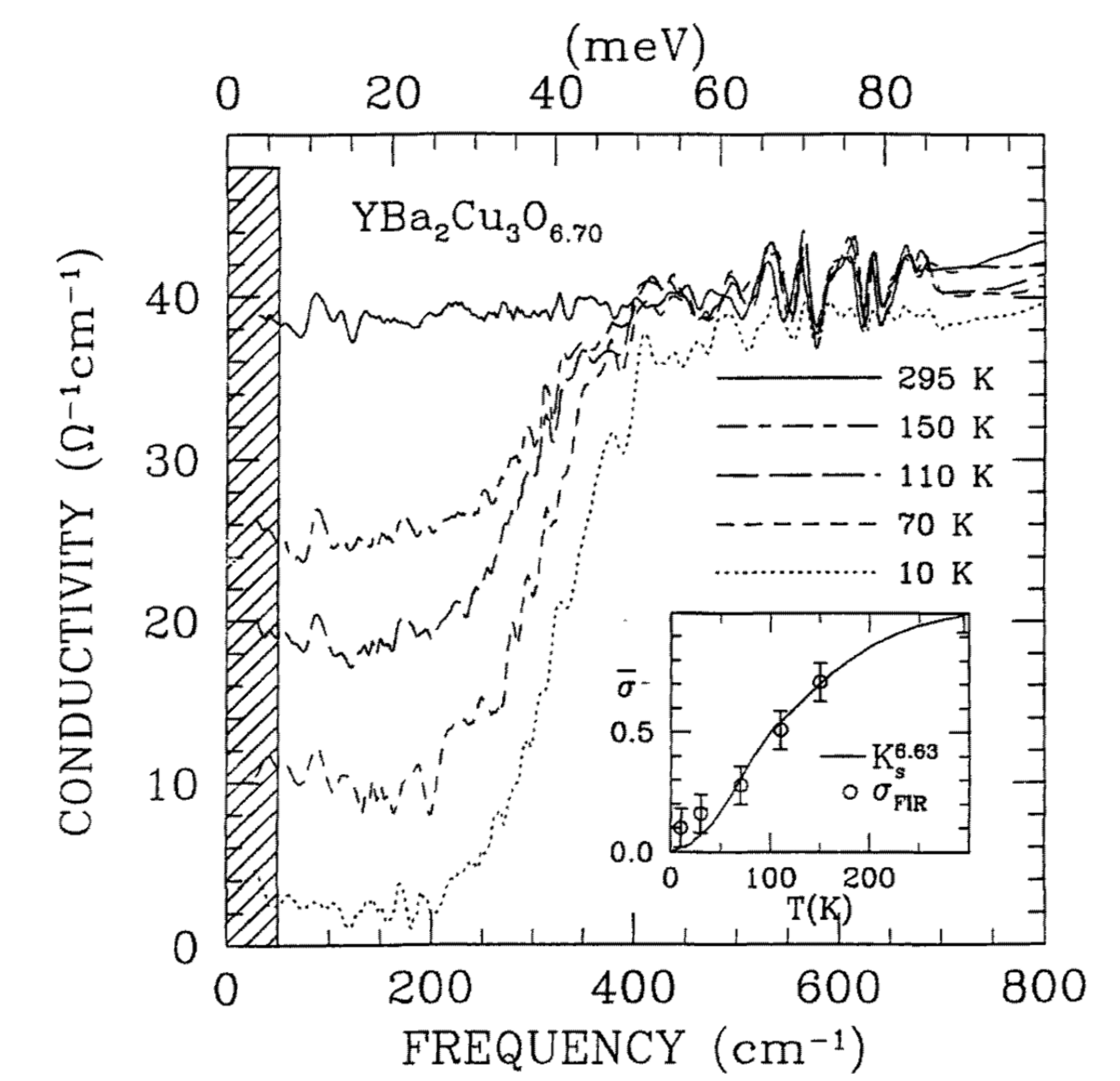}
\caption{C-axis Infrared conductivity. Fig. 2 of Ref.~\cite{puchkov1996pseudogap}}
\label{Fig:  c axis data}
\end{center}
\end{figure}

\begin{figure}[htb]
\begin{center}
\includegraphics[width=0.95\linewidth]{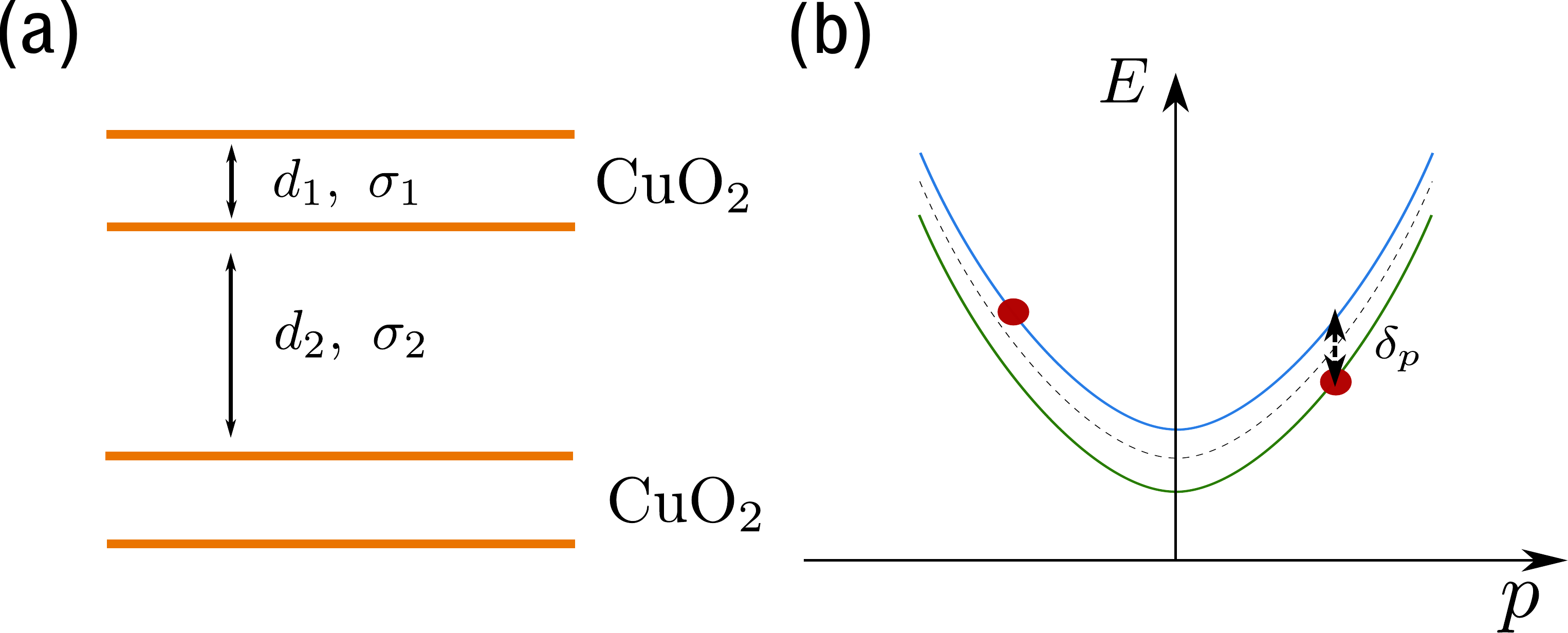}
\caption{(a) Sketch of bilayer cuprates (for example, YBCO, Bi2212). Each orange line represents one $\text{CuO}_2$ layer. (b) Split boson bands due to interlayer hopping. AC voltage between two layers can excite a boson and a vacancy of boson (red dots).}
\label{Fig:  c axis infrared}
\end{center}
\end{figure}

Bilayer cuprates, such as YBCO, consists of two $\text{CuO}_2$ layers separated by only one atomic layer ($d_1$ in Fig.~\ref{Fig:  c axis infrared}, each orange line represents a $\text{CuO}_2$ layer). The distance from these two layers to neighboring bilayers is 3 or 4 atomic spacing ($d_2$ in Fig.~\ref{Fig:  c axis infrared}). The tunneling conductivity between the small barrier ($\s_1$) and the large barrier ($\s_2$) can be calculated perturbatively in the corresponding interlayer hopping. We expect $|\s_1|\gg|\s_2|$, since the hopping decays exponentially. However, the c-axis conductivity, measured by reflectance of light with electric field polarizing in c-axis is mixture of the two tunneling conductivity. By effective medium approximation,

\bea
\e = \frac{d_1+d_2}{d_1/\e_1 +d_2/\e_2},\\\text{ where }\e_i = 1+\frac{i\s_i}{\omega\e_0}, \e = 1+\frac{i\s}{\omega\e_0}.
\eea

When $\s_1\gg\s_2$, there are two limiting possibilities. The first one is that $\s_1\sim \omega\e_0,\ \e_1\sim 1$, and we can ignore $\s_2$. In this case, $\s\sim \omega\epsilon_0, \e\sim 1$, and the measured c-axis conductivity reflects tunneling between the small barrier. The second possibility is that $\s_1\gg\omega\e_0,\ \e_1\gg 1,\e_2$. In this case, we can ignore $d_1/\e_1$, hence $\e = \e_2d/d_2$. Note that theoretically, $\e$ can be much larger than 1 only in the second case, under the condition $\e_1\gg\e_2\gg1$. Experimentally, in the infrared region, $\omega\sim 40$meV, $\text{Re}\,\s \sim 40\,\Omega^{-1}\text{cm}^{-1}$, $\text{Re}\,\s/(\omega\epsilon_0) \sim 40$ (See Fig.~\ref{Fig:  c axis data} and Ref.~\cite{puchkov1996pseudogap,PhysRevB.43.7942}). Away from sharp resonances, we are clearly in the second limit. Thus the measured c-axis conductivity in bilayer cuprates reflects tunneling between the larger barrier ($d_2$ in Fig.~\ref{Fig:  c axis infrared}). Intuitively, the smaller barrier is so conductive that the majority of voltage drop is on the larger barrier, which contribute most to the measured conductivity.

Now we analyze boson contribution to the c-axis conductivity. Across the large barrier, we expect the interlayer hopping of boson to be considerably smaller than the interlayer hopping of fermion. Nonetheless, it can still contribute. We use the following phenomenological model for the coupling between two layers across the large barrier.
\bea
\mathcal{L} = \sum_{i =1,2}\frac{1}{2}|(\partial_{\mu} + ie^{*}A_{\mu})\phi_i|^2 - \frac{1}{2}|\D_b|^2|\phi_i|^2 -\nonumber\\ \frac{1}{2}g(\phi_1^*\phi_2 + \phi^*_2\phi_1),
\eea
where $\phi_1$ and $\phi_2$ are the boson fields in the upper and lower layer. Define $\phi_{\pm}\equiv (\phi_1\pm\phi_2)/\sqrt{2}$. Canonical quantization gives $\phi_{\pm} = \frac{1}{E_{p,\pm}}(a_{p,\pm} + b_{-p,\pm}^{\dagger})$, where $E_{p,\pm} = \sqrt{p^2+\D_b^2\pm g_p}\equiv E_p\pm\delta_p$. For small $g$, $E_p\simeq \sqrt{p^2+\D_b^2}$, $\delta_p\simeq g_p/E_p$. Note that we have set the in-plane velocity $v_b=1$ for convenience. The momentum dependence of $g$ comes from the tunneling matrix elements on the lattice scale. Electric field couples to the density difference of the two layers: $\delta H = \frac{1}{2}(\rho_1-\rho_2)Ed_2$, and

\bea
\rho_1 - \rho_2 &=& \frac{ie^*}{2}(\partial_t\phi_1^*\cdot\phi_1 - \phi_1^*\partial_t\phi_1 -\partial_t\phi_2^*\cdot\phi_2 + \phi_2^*\partial_t\phi_2)\ \\
&=& \frac{ie^*}{2} (\partial_t\phi_+^*\cdot\phi_- + \partial_t\phi_-^*\cdot\phi_+ - \phi_+^*\partial_t\phi_- - \phi_-^*\partial_t\phi_+)\nonumber\\
&=& \frac{e^*}{2}(\frac{\sqrt{E_{p,+}}}{\sqrt{E_{p,-}}} - \frac{\sqrt{E_{p,-}}}{\sqrt{E_{p,+}}})(a_{p,-}^{\dagger}b_{-p,+}^{\dagger} - a_{p,+}^{\dagger}b_{-p,-}^{\dagger})\nonumber\\
& & + \dots\nonumber\\
&\simeq& \frac{e^*\delta_p}{2E_p}(a_{p,-}^{\dagger}b_{-p,+}^{\dagger} - a_{p,+}^{\dagger}b_{-p,-}^{\dagger}) + \dots
\eea
where $\dots$ represents terms that annihilate the ground state. Electric field can excite a pair of bosons with opposite charge, one to the plus band and one to the minus band, as illustrated in Fig.~\ref{Fig:  c axis infrared}(a). The current between the two layers is $j = \partial_t \rho_1 = \frac{1}{2}\partial_t(\rho_1 - \rho_2)$. By Kubo formula, the c-axis conductivity is

\bea
\text{Re}\,\sigma_{2} &=& \pi d_2\omega\sum_n|\<n|\frac{\rho_1-\rho_2}{2}|0\>|^2\ \delta(\omega - E_n + E_0)\ \ \ \  \\
&=& 2\pi\omega d\int \dbar^2p \frac{(e^*)^2\delta_p^2}{16E_p^2}\ \delta(\omega - E_{p,+} - E_{p,-})\\
&=& \frac{e^2}{4\hbar}\frac{\delta_\omega^2 d_2}{v_b^2\hbar^2}\ \theta(\omega - 2\D_b)
\eea
In the last line, we restore $\hbar$ and the boson velocity $v_b$, which we previously set to 1. $\delta_\omega$ is the energy splitting between the two excited bosons at frequency $\omega$, the frequency dependence comes from the momentum dependence of $\delta_p$ (see Fig.~\ref{Fig:  c axis infrared}(b)). The conductivity has a step-function onset because of the step-function onset of density of states in 2D. This behavior matches the measured c-axis conductivity. However, the boson contribution is proportional to $\delta_p^2$, which is the forth power of single-electron tunneling. On the other hand, fermion tunneling also gives a step-function contribution to c-axis conductivity. Since fermion interlayer hopping is considerably larger than boson interlayer hopping, we expect that a considerable part of the c-axis conductivity comes from small-gap fermions in the fluctuating PDW bands.

\section{Finite-size extrapolation of boson and fermion gap}\label{Appendix: DMRG}

\begin{figure}[h]
\begin{center}
\includegraphics[width=0.6\linewidth]{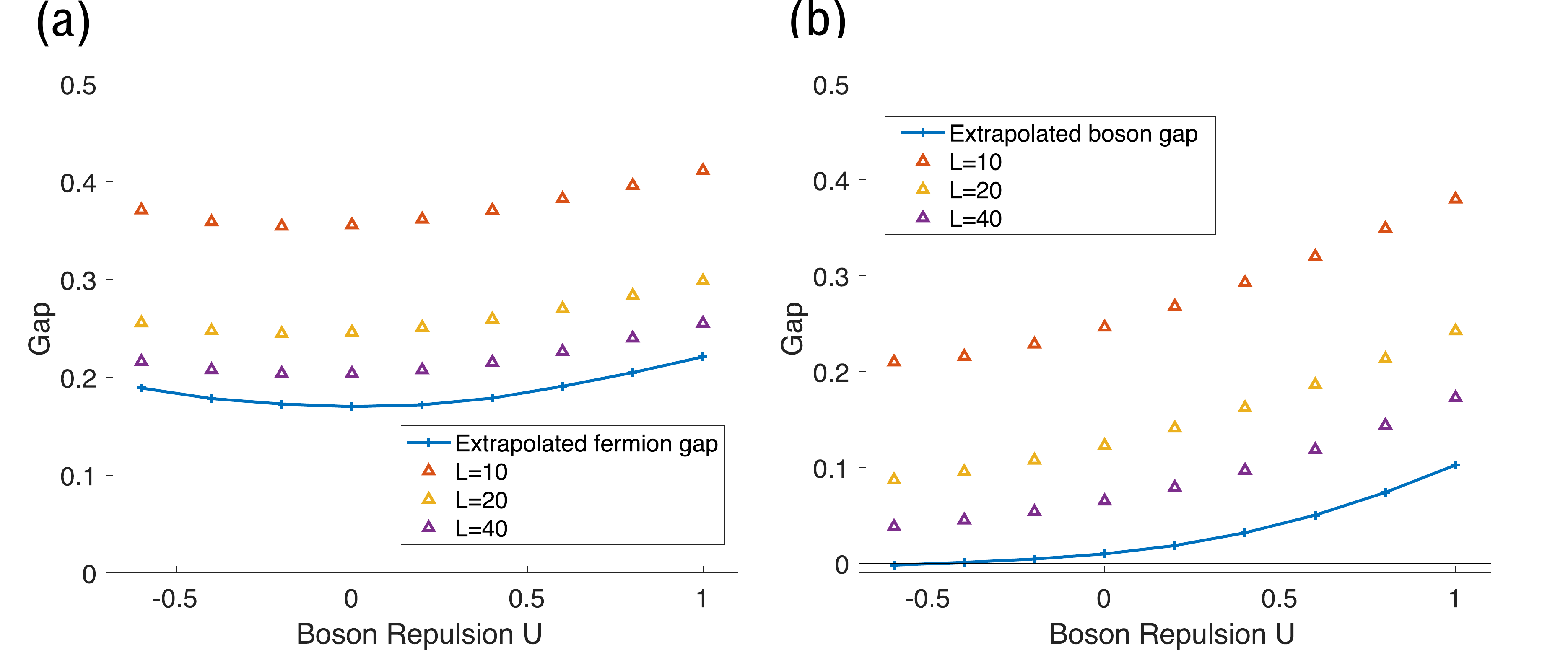}
\caption{(a) Extrapolation of fermion gap, $t=1.0, p= 0.5$. (b) Extrapolation of boson gap, $t=1.0, p= 0.5$.}
\label{Fig: DMRG extrapolation p05}
\end{center}
\end{figure}

We compute boson gaps and fermion gaps of the 1D model in Sec.~\ref{subsection: 1D numerics} (as a function of the boson repulsion $U$) on system with length $L=10,20,40$, and then fit the gap to the form

\bea
E(L) = E_{\infty} + a/L + b/L^2
\eea
to get the thermodynamic gap $E_{\infty}$. Fig.~\ref{Fig: DMRG extrapolation p05} shows finite-size gaps together with extrapolated gaps for $p=0.5$.

\onecolumngrid

\twocolumngrid

\end{document}